\documentclass[fleqn,usenatbib]{mnras}
\usepackage[normalem]{ulem} 
\usepackage{newtxtext,newtxmath}
\usepackage{ulem}

\usepackage[T1]{fontenc}
\usepackage{ae,aecompl}
\usepackage{comment}

\usepackage{stackengine,graphicx}	
\usepackage{amsmath}	
\usepackage{amssymb}	

\usepackage{multirow} 
\usepackage{multicol}

\newcommand{\sgra}{Sgr~A$^{\star}$}
\newcommand{\solM}{$M_{\odot}$}
\newcommand{\ka}{$K_{\alpha}$}
\newcommand{\apec}{\textit{APEC}}
\newcommand{\nus}{\textit{NuSTAR}}
\newcommand{\xmm}{\textit{XMM--Newton}}
\newcommand{\chan}{\textit{Chandra}}

\title[Investigating of the Arches cluster non-thermal emission]{Investigating the origin of the faint non-thermal emission of the Arches cluster using the 2015-2016 \nus\ and \xmm\ X-ray observations}

\author[Kuznetsova et al.]{Ekaterina Kuznetsova,$^{1}$\thanks{E-mail: eakuznetsova@cosmos.ru}
Roman Krivonos$,^{1}$ Ma{\"i}ca Clavel$,^{2}$ Alexander Lutovinov$,^{1, 3}$ \and
Dmitry Chernyshov$,^{4}$ JaeSub Hong$,^{5}$ Kaya Mori$,^{6}$ Gabriele Ponti$,^{7}$ John Tomsick$,^{8}$ \and Shuo Zhang$^{9}$
\\
$^{1}$Space Research Institute of the Russian Academy of Sciences, Profsoyuznaya 84/32, 117997 Moscow, Russia\\
$^{2}$Univ. Grenoble Alpes, CNRS, IPAG, F38000 Grenoble, France\\
$^{3}$Higher School of Economics, Myasnitskaya 20, 101000 Moscow, Russia\\
$^{4}$I.E. Tamm Theoretical Physics Division of P.N. Lebedev Institute of Physics, Leninskii pr. 53, 119991 Moscow, Russia\\
$^{5}$Harvard-Smithsonian Center for Astrophysics, 60 Garden St., Cambridge, MA 02138, USA\\
$^{6}$Columbia Astrophysics Laboratory, Columbia University, New York, NY 10027, USA\\
$^{7}$Max-Planck-Institut f\"{u}r extraterrestrische Physik, Giessenbachstrasse 1, Garching, 85748, Germany\\
$^{8}$Space Sciences Laboratory, University of California, 7 Gauss Way, Berkeley, CA 94720-7450, USA\\
$^{9}$MIT Kavli Institute, 77 Massachusetts Ave, Cambridge, MA 02139 USA\\
}

\date{Accepted XXX. Received YYY; in original form ZZZ}
\pubyear{2018}

\begin{document}
\label{firstpage}
\pagerange{\pageref{firstpage}--\pageref{lastpage}}
\maketitle


\begin{abstract}
Recent \nus\ and \xmm\ observations of the molecular cloud around the Arches stellar cluster demonstrate a dramatic change both in morphology and intensity of its non-thermal X-ray emission, similar to that observed in many molecular clouds of the Central Molecular Zone at the Galactic Center. These variations trace the propagation of illuminating fronts, presumably induced by past flaring activities of \sgra. In this paper we present results of a long \nus\ observation of the Arches complex in 2016, taken a year after the previous \textit{XMM}$+$\nus\ observations which revealed a strong decline in the cloud emission. The 2016 \nus\ observation shows that both the non-thermal continuum emission and the Fe \ka\ 6.4~keV line flux are consistent with the level measured in 2015. No significant variation has been detected in both spectral shape and Fe \ka\ equivalent width $EW_{\rm 6.4\ keV}$, which may be interpreted as the intensity of the Arches non-thermal emission reaching its stationary level. At the same time, the measured 2016 non-thermal flux is not formally in disagreement with the declining trend observed in 2007-2015. Thus, we cannot assess whether the non-thermal emission has reached a stationary level in 2016, and new observations, separated by a longer time period, are needed to draw stringent conclusions. Detailed spectral analysis of three bright clumps of the Arches molecular cloud performed for the first time showed different $EW_{\rm 6.4\ keV}$ and absorption. This is a strong hint that the X-ray emission from the molecular cloud is a mix of two components with different origins.
\end{abstract}



\section{Introduction}

The Arches cluster \citep{cotera96, serabyn98} is a massive star cluster, which is located in the Galactic Center (GC) region at the projected angular distance of $28$~pc from \sgra. The cluster contains more than 160 O-type stars with initial masses larger than 20~\solM\ and its average mass density is $\sim3\times10^5\ M_{\odot}{\rm pc}^{-3}$ \citep{figer99, figer02}. The Arches core is about $9''$ ($\sim0.35$~pc at 8~kpc) in radius \citep{figer99}. 

The first X-ray observation of the region around the Arches cluster with \chan\ \citep{zadeh} revealed its bright X-ray emission. This emission, characterized by the 6.7~keV Fe line, was detected in the narrow region consistent with the Arches cluster position (hereafter ``the core''). The thermal emission of the cluster was attributed to the multiple collisions between strong winds of massive stars \citep{zadeh, wang, capellia}. In addition to the cluster's core emission, the extended emission of a different origin, spatially consistent with a nearby molecular cloud, was detected around the cluster (hereafter ``the cloud''). The observed prominent Fe \ka\ line and the power-law continuum emission point out to its non-thermal nature. \citep{wang, tsujimoto, capellib, T12, K14, K17, clavel14}.

Two main hypotheses have been developed to explain the non-thermal emission from the Arches ``cloud'' emission. The first approach implies irradiation of the molecular cloud by an external X-ray source. The possible X-ray sources inside or nearby the cloud have been excluded, based on the required X-ray luminosity \citep{capellib} to explain the 6.4 keV line flux \citep{T12, K14}. The sufficient luminosity could be provided by a nearby X-ray source, e.g. 1E1740.7--2942 \citep{1993ApJ...407..752C} or with the past activity of \sgra, as originally suggested by \cite{1993ApJ...407..606S} to explain the fluorescent line emission observed in the giant molecular cloud Sgr~B2 in the GC region \citep[see also][]{1996PASJ...48..249K, 1998MNRAS.297.1279S, 2000ApJ...534..283M, 2004A&A...425L..49R, 2010ApJ...719..143T, 2015ApJ...815..132Z}. The hypothesis of the past \sgra's flaring activity is also supported by the discovery of a propagation of Fe K echos in the Central Molecular Zone (CMZ), where \sgra\ is presumably the source of the illumination on the clouds \citep{2010ApJ...714..732P,2013ASSP...34..331P,clavel13,2013PASJ...65...33R,2017MNRAS.465...45C,2017MNRAS.471.3293C,2018A&A...612A.102T}. 

Alternatively, a fluorescent emission of the molecular cloud can be a result of bombardment by low-energy cosmic ray (CR) particles \citep{capellib, T12}. The steady non-thermal continuum and Fe~K$\alpha$ flux observed in the Arches cloud for a decade was initially considered as a strong evidence for the CR heating of the Arches molecular cloud \citep{T12}. However, a recently discovered time variability of the non-thermal power-law continuum and 6.4~keV line \citep{clavel14}, strongly contradicts the CR scenario \citep[see e.g.][]{2014APh....54...33D} for a significant fraction of this emission. 

Using the 2015 \textit{XMM}$+$\nus\ data set, \cite{K17} showed that the Arches cloud experienced a strong emission decrease, along with a significant decrease of the Fe \ka\ equivalent width (EW). This could indicate either a change in the reflection geometry (illumination of different structures along the line of sight) or that the putative CR component has become more dominant. In any case, the question remains whether the non-thermal continuum and the 6.4~keV line emission will decrease to a zero flux level when the X-ray illuminating front leaves the molecular cloud or if some background CR heating takes place.

As discussed above, the variable component of the Arches complex non-thermal emission is generated by Thomson scattering of X-ray flare photons leaving the Arches cloud complex. \cite{2018ApJ...863...85C} suggested that the other (possibly stationary) component can be caused by either i) Thompson scattering of photons from the same or a different X-ray flare, ii)  another molecular cloud complex which is situated at large distance from the Arches cluster and on the same line of sight, or iii) the excitation of CR particles, however more restrictions must be satisfied. The authors argue that in case of a single cloud complex (i) the $EW$ variations are likely associated with iron abundance changes in it during the flare moving across the cloud. For the second case with two different clouds (ii), \cite{2018ApJ...863...85C} calculated the temporal delay of the Compton echo for the first variable component at about 100 yrs. This is in agreement with \citet{2017MNRAS.465...45C} who estimated X-ray flare event with 110 yrs age. For the second variable component the estimated time delay is about 230 yrs \citep{2018ApJ...863...85C}. It is more likely that these two components were caused by the different flares rather than the same flare. The suggestion of two different flares are consistent with scenario of two flares of \sgra\ 110 and 240 yrs ago \citep{clavel13,2018A&A...612A.102T,chuard18}.

Changes in the morphology of the Arches cluster cloud are similar to those seen for Sgr~B2, where the non-thermal emission likely reaching the background level as well, revealing a substructure of two compact cores and a newly emerging cloud that confirms the propagation of the illuminating front(s) from an X-ray flare \citep{2015ApJ...815..132Z}. These similarities between the Arches cloud emission and GC molecular clouds \citep{K17} support the reflection mechanism of the non-thermal emission.

In this work we analyzed observations of the Arches cluster region with \nus\ in 2015-2016 and \xmm\ in 2015, with the aim to monitor the non-thermal emission around the cluster. Additionally, we carried out detailed spectral analysis of the emission clumps detected with \xmm\ observations in 2015. The paper is structured as follows: in Sect.~\ref{sec:obs} we describe observations of the Arches cluster with \nus\ in 2016 and outline the data analysis; 2016 \nus\ data analysis morphology and spectral analyses are presented in Sect.~\ref{sec:2016}; Sect.~\ref{sec:15-16} contains the analysis of the combined \nus\ and \xmm\ 2015-2016 observations. We investigate the substructure of the Arches cloud non-thermal emission revealed by the 6.4~keV line emission clumps observed with \xmm\ (Sect.~\ref{sec:clumps}). We discuss and summarize the obtained results in Sect.~\ref{sec:summ}.

\section{Observation and data analysis}
\label{sec:obs}

The Arches cluster region was observed in October 2016 during the \nus\ Legacy survey\footnote{\url{https://www.nustar.caltech.edu/page/legacy_surveys}} of the Sgr~A molecular clouds with a total exposure time of 150~ks. The list of the used \nus\ observations is shown in Table~\ref{tab:obs}.

\begin{figure}
\center 
\includegraphics[width=1\columnwidth]{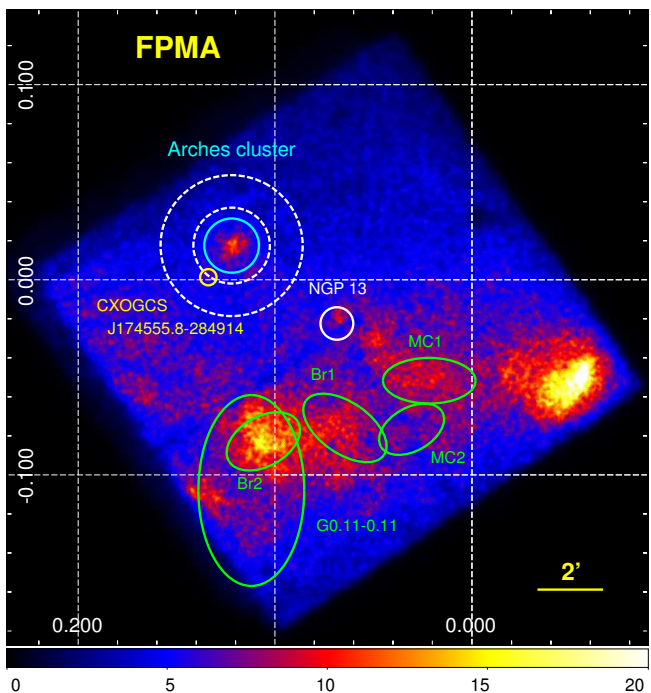} 
\caption{ \nus\ $3-79$~keV image in Galactic coordinates of Sgr~A molecular clouds region, including the Arches cluster, taken in 2016 (FPMA data only, exposure 150~ks). The extraction of the source and background spectra was done, respectively, in the circle $R=50''$ and annulus ($70''<R<130''$). Solid elliptical sky regions shows positions of selected Sgr~A molecular clouds \citep{clavel13}. NGP 13 label denotes position of a hard X-ray source detected in the \nus\ survey of the GC region \citep{2016ApJ...825..132H}. Small solid circle labeled CXOGCS J174555.8--284914 shows the position of corresponding X-ray source (Sect.~\ref{subsec:15-16_2D}). The image was smoothed using {\sc dmimgadapt} task from {\sc ciao-v.4.9} with the following parameters: tophat kernel, smoothing scales $1-20$, the number of scales is 30, the minimum number of counts is 20.}
\label{fig:fpma}
\end{figure}

\begin{table}
\noindent
\centering
\caption{The list of the Arches cluster observations, which have been used in this work.}
\label{tab:obs}
\centering
\vspace{1mm}
\begin{tabular}{c|c|c|c}
\hline \hline
Experiment & ObsID & Start Date & Exposure, s \\
\hline
\textit{XMM} & 0762250301 & 2015-09-27 15:48:39 & 114000 \\
\hline
\nus\  & 40101001002 & 2015-10-19 06:21:08 & 107189 \\
      & 40101001004 & 2015-10-25 13:56:08 & 107856 \\
      & 40202001002 & 2016-10-28 13:16:08 & 150856 \\
\hline
\end{tabular}\\
\vspace{3mm}
\end{table}

The hard X-ray orbital telescope \nus\ \citep{harrison}, launched in 2012, provides imaging at energies above 10~keV with sub-arcminute angular resolution. \nus\ carries two identical co-aligned X-ray telescope modules (referred as FPMA and FPMB) with a full width at half maximum (FWHM) of $18''$. The energy response covers a wide range from 3 to 79~keV with a spectral resolution of 400~eV (FWHM) at 10~keV.

Due to \nus's known `stray-light' issue, when detectors can be illuminated by X-rays passing outside the field of view the X-ray optics \citep{straylight}, the data from FPMB are strongly contaminated by photons from the bright X-ray source GX~3$+$1, which forced us to exclude FPMB data in this work. Fig.~\ref{fig:fpma} shows the FPMA image of the Arches cluster region in the full $3-79$~keV energy band. Arches was observed at $\sim4'$ distance from the optical axis, which led to a somewhat efficiency decrease, however without a visible distortion of the point spread function (PSF).

It is worth noting that celestial coordinates of each photon registered by \nus\ are subject to a systematic offset, which can be as high as $14''$ \citep{2015ApJ...814...94M}. We noticed an offset between the cluster's centroid position and its catalogued coordinates at the level of $3-4''$. Similar to \cite{K17}, we performed an astrometric correction by shifting coordinates of each photon using a reference position of the Arches core centroid, measured in the first \nus\ observation of the Arches cluster in 2012 \citep{K14}, when the offset was negligible.

We extracted spectrum of the Arches cluster complex, and corresponding response matrices, using the \textit{nuproducts} task, a part of \nus\ Data Analysis Software package ({\sc nustardas v.1.8.0}), built in {\sc heasoft} software (version 6.22). Note that in this paper we often compare our results with those of \cite{K17} who used older {\sc heasoft} version 6.17. We checked that different versions do not introduce strong deviations to the results. The source extraction region was a circle with a radius of $50''$ positioned at the cluster position $R.A.=17^h45^m50.52^s$, $Dec.=-28^{\circ}49'22.41''$ \citep{K14,K17}. The background spectrum was obtained from the annulus region within the radii range $R=70-130''$ (see Fig.~\ref{fig:fpma}), excluding $R=15''$ circular region of the source CXOGCS J174555.8--284914 \citep{J174555}, as described in Sect.~\ref{subsec:15-16_2D}. The interactive spectral and 2D image analysis were done using {\sc xspec} \citep{xspec} and {\sc sherpa} \citep{sherpa} tools included, respectively, in {\sc heasoft 6.22} and {\sc ciao-v.4.9} \citep{ciao} software packages.

\section{\nus\ observations of the Arches cluster in 2016}
\label{sec:2016}

\subsection{2D image analysis}
\label{subsec:2016_2D}

Since the Arches cluster demonstrated a strong drop of its non-thermal emission in 2007-2015 \citep{clavel14,K17}, one would expect a further fall even below the detection threshold. The main question we address in this section is whether the Arches stellar cluster is still surrounded by the spatially extended emission, previously detected with \nus\ in 2012-2015 \citep{K14,K17}. The \nus\ angular resolution of $18''$ (FWHM) does not allow us to directly resolved the emission around the cluster's core. Therefore we use a simple 2D image analysis with the {\sc sherpa} package to estimate a spatial extension of the cluster emission compared to that expected from the \nus\ PSF. {\sc sherpa} performs spatial fitting of a data image using a 2D source model convolved with supplied instrumental PSF.

We first analyzed the 2016 \nus\ $3-79$~keV image of the Arches cluster region using 2D Gaussian model \texttt{G1} with a fixed $4''$ FWHM (PSF smearing effect) and free center position, representing a point-like X-ray source. The model is represented in the {\sc sherpa} notation as \texttt{psf(gauss2d.G1)*emap+const2d.bkg*emap}, where \texttt{G1} is 2D Gaussian, associated with the emission of the stellar cluster, \texttt{emap} contains exposure map, and \texttt{const2d.bkg} component estimates the background count rate, in the assumption of a flat distribution. As shown in Fig.~\ref{fig:fpma}, the fitting area was limited by a circle with $R=50''$ around the cluster's centroid position and the background count rate was estimated in the annulus $70''<R<130''$ with the same center position, excluding the circle region with $R=15''$ of X-ray source CXOGCS~J174555.8--284914 (Sect.~\ref{subsec:15-16_2D}). The difference between the $3-79$~keV image and the best-fit model is characterized by strong deviations indicating that the X-ray emission of the Arches cluster is not consistent with a point-like X-ray source \citep{wang,J174555}. To quantitatively estimate the spatial scale of the extended emission of the Arches cluster we added a second 2D Gaussian model \texttt{G2} with a free position and FWHM parameter. The list of free  parameters of this new complex model includes the position of \texttt{G1}, the position and width of \texttt{G2}, and the flat background level. The extended model provides a good fit to the data, characterized by no strong deviations in the residuals, and fit statistics $\chi_{\rm r}^{2}\approx1$. The fitting procedure estimates a spatial scale of the extended emission \texttt{G2} as being $32''\pm7''$ FWHM, which is significantly larger than the \nus\ PSF (18$''$ FWHM). This result indicates that the spatially extended emission of the Arches cluster is still present in 2016. The position of the \texttt{G2} extended component is shifted off by $13''\pm3''$ from the position of the Arches stellar core (\texttt{G1}), which also indicates that the extended emission is not associated with the stellar cluster.

\subsection{Spectral analysis}
\label{subsec:2016spec}

As shown in the previous subsection, the extended emission, associated with the non-thermal radiation of the neutral or low ionization state material of the molecular cloud is still present around the stellar cluster. For quantitative evaluation, we perform a spectral analysis comparing the recent 2016 \nus\ observations with previous results from by \cite{K17} based on the \nus\ and \xmm\ data acquired in 2015.

\begin{figure}
\center
\includegraphics[width=1\columnwidth]{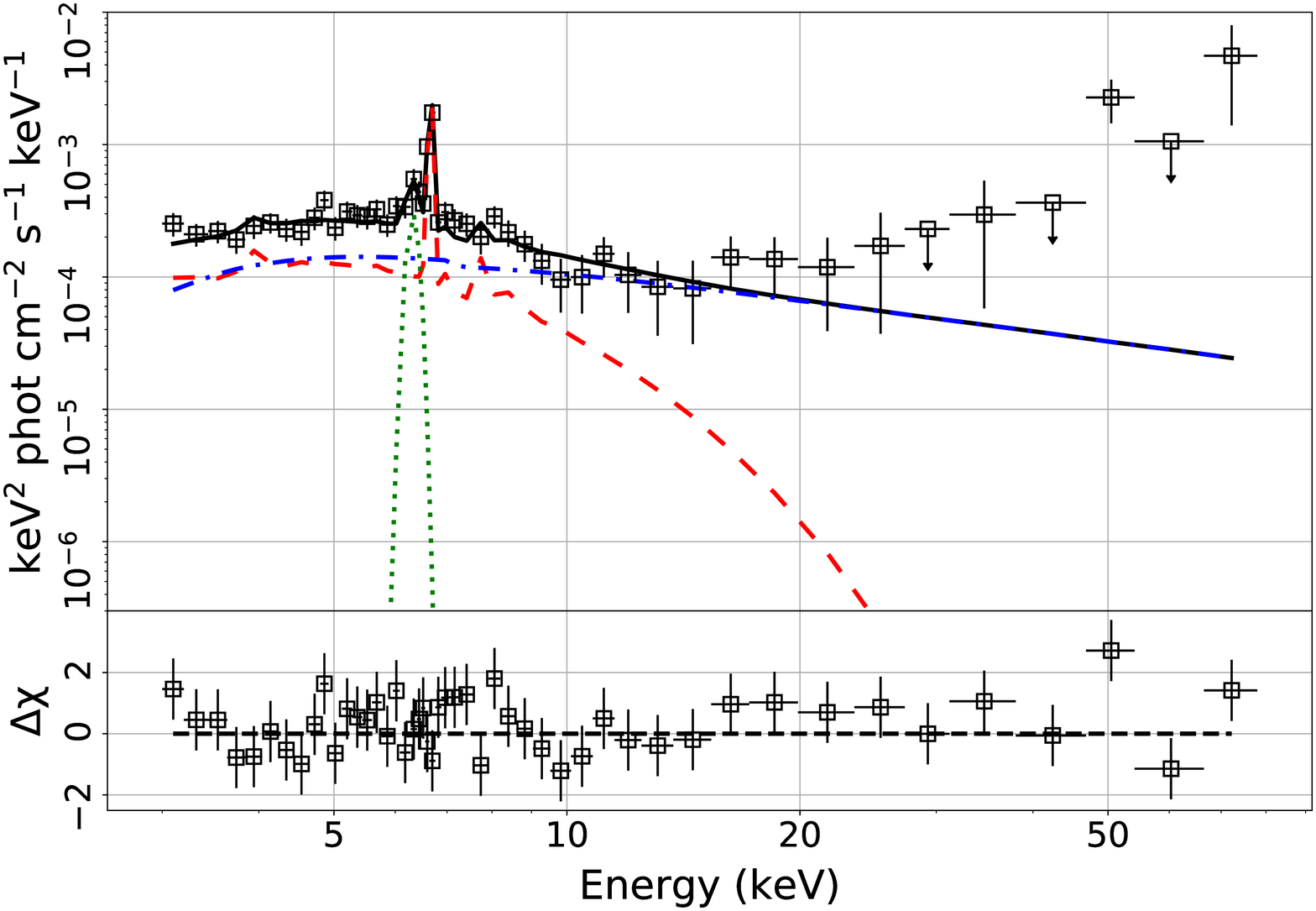}
\caption{The Arches cluster \nus\ 2016 spectrum extracted from the $50''$ circle region around the cluster (detector module A only).  Solid, dashed, dotted and dash-dotted lines are the total model (see Table~\ref{tab:spec}), thermal \apec, Fe~\ka\ line and non-thermal power-law components respectively.}
\label{fig:spec}
\end{figure}

We reduced the 2016 \nus\ observation of the Arches cluster  and extracted source and background spectra from $50''$ circular and $70-130''$ annulus regions, respectively, as described in Section~\ref{sec:obs}. The background-subtracted FPMA spectrum of the Arches cluster region is shown in Fig.~\ref{fig:spec}. To fit the spectrum, we applied the same spectral model $wabs\times(APEC+Gaussian+powerlaw)$ used in \cite{K17}, which contains the thermal emission of the stellar cluster described by the collisionally ionized plasma emission model (\apec) and the non-thermal emission of the molecular cloud in the form of power-law and Gaussian 6.4~keV line. The thermal emission is characterized by $kT$, Z/Z$_{\odot}$, and $I_{\rm kT}$ parameters  describing, respectively, the temperature in keV, the metallicity relative to the solar one, and the normalization in units of $10^{-18}\int n_{\rm e}n_{\rm H}dV/(4\pi D^{2})$, where $n_{\rm e}$ and $n_{\rm H}$ are the electron and proton number densities in units of $\textrm{cm}^{-3}$, and D is the cluster distance in cm. The metallicity was fixed at $Z=1.7Z_{\odot}$ according to \cite{T12}. The non-thermal emission is approximated by a power-law continuum and Fe 6.4~keV line, represented by a Gaussian with the width fixed at 0.1~keV. All the emission spectral components are subject to interstellar photo-electric absorption modeled with \textit{wabs} in {\sc xspec}. Due to limited energy response of the \nus\ below 5~keV, the fitting procedure does not allow us to constrain the absorption, and we fixed it at $N_{\rm H}=9.5\times10^{22}$~cm$^{-2}$ \citep{T12}. The equivalent width (EW) of the 6.4~keV line was estimated with the {\sc sherpa} package with respect to the \textit{powerlaw} component only. The fitting procedure demonstrated a strong correlation between components, causing a degeneracy of model parameters: plasma temperature $kT$ and power-law photon index $\Gamma$. To overcome this, for the following analysis we fixed the \apec\ temperature at $2.4$~keV, as measured by \cite{K17} in 2015 \nus\ observations. Best-fitting model parameters are listed in Table~\ref{tab:spec}. For comparison with 2015 \nus\ observations we included to the table the best-fitting parameters of the same model adopted from \cite{K17}. 

The spectral analysis of the \nus\ observations of the Arches cluster complex in 2015 and 2016 does not demonstrate strong variations of both thermal emission of the cluster and non-thermal emission of the surrounding molecular cloud. The main parameters of the non-thermal emission, power-law slope and normalization, and $EW$ of the iron 6.4~keV line, are consistent within uncertainties between the observations. We conclude that further observations of the Arches cluster region are needed to confirm or deny the stationary level of the Arches cloud non-thermal emission.

\begin{table*}
\noindent
\centering
\caption{Best-fit model parameters for the Arches cluster region, measured with {\nus} in 2015 \citep{K17} and 2016 (this work). Model: $wabs\times(APEC + Gaussian + power law)$.}
\label{tab:spec}
\centering
\vspace{1mm}
 \begin{tabular}{c|c|c|c|c}
\hline\hline
Parameter & Unit & 2015 yr  & 2016 yr & $2015 \& 2016$ \\
\hline
$N_{\rm H}$ & $10^{22}$~cm$^{-2}$ & $9.5$ (fixed) &  $9.5$ (fixed) & $9.5$ (fixed) \\ 
\hline
$kT$ & keV & $2.4_{-0.4}^{+1.5}$ &  $2.4$ (fixed) & $2.4$ (fixed) \\
$I_{\rm kT}$ & see Sect. \ref{subsec:2016spec} & $9.2\pm5.3$ & $8.4_{-4.4}^{+2.1}$ & $9.1_{-1.4}^{+1.2}$ \\
\hline
$\Delta E_{\rm 6.4\ keV}$ & keV &  0.1 (fixed) & 0.1 (fixed) & 0.1 (fixed) \\
$E_{\rm 6.4\ keV}$ & keV &  $6.3\pm1.1$ & $6.3\pm0.3$ &  $6.3\pm0.1$\\
$F_{\rm 6.4\ keV}$ & $10^{-6}$ ph~cm$^{-2}$~s$^{-1}$ &  $2.6\pm1.5$ & $2.5_{-1.7}^{+3.1}$ & $2.4\pm1.2$ \\
$EW_{\rm 6.4\ keV}$ & eV & $450\pm150$ &  $580\pm250$ &  $540\pm140$ \\
\hline
$\Gamma$& & $2.4_{-0.3}^{+0.6}$ & $2.7_{-0.5}^{+0.4}$ & $2.6\pm0.2$ \\
$F^{\rm pow}_{\rm 3-20\ keV}$& 10$^{-13}$~erg~cm$^{-2}$~s$^{-1}$ &
$5.6_{-0.9}^{+1.2}$  & $4.4_{-3.1}^{+2.1}$ & $5.8\pm0.8$\\
\hline
$\chi^{2}_{\rm r}$/d.o.f.& & 0.97/364  &  0.82/204 &  0.91/574  \\
\hline
\end{tabular}\\
\vspace{3mm}
\end{table*}

\section{The combined 2015-2016 data set}
\label{sec:15-16}
\subsection{2D image analysis}
\label{subsec:15-16_2D}

\begin{figure*}
\center
\includegraphics[width=0.32\textwidth]{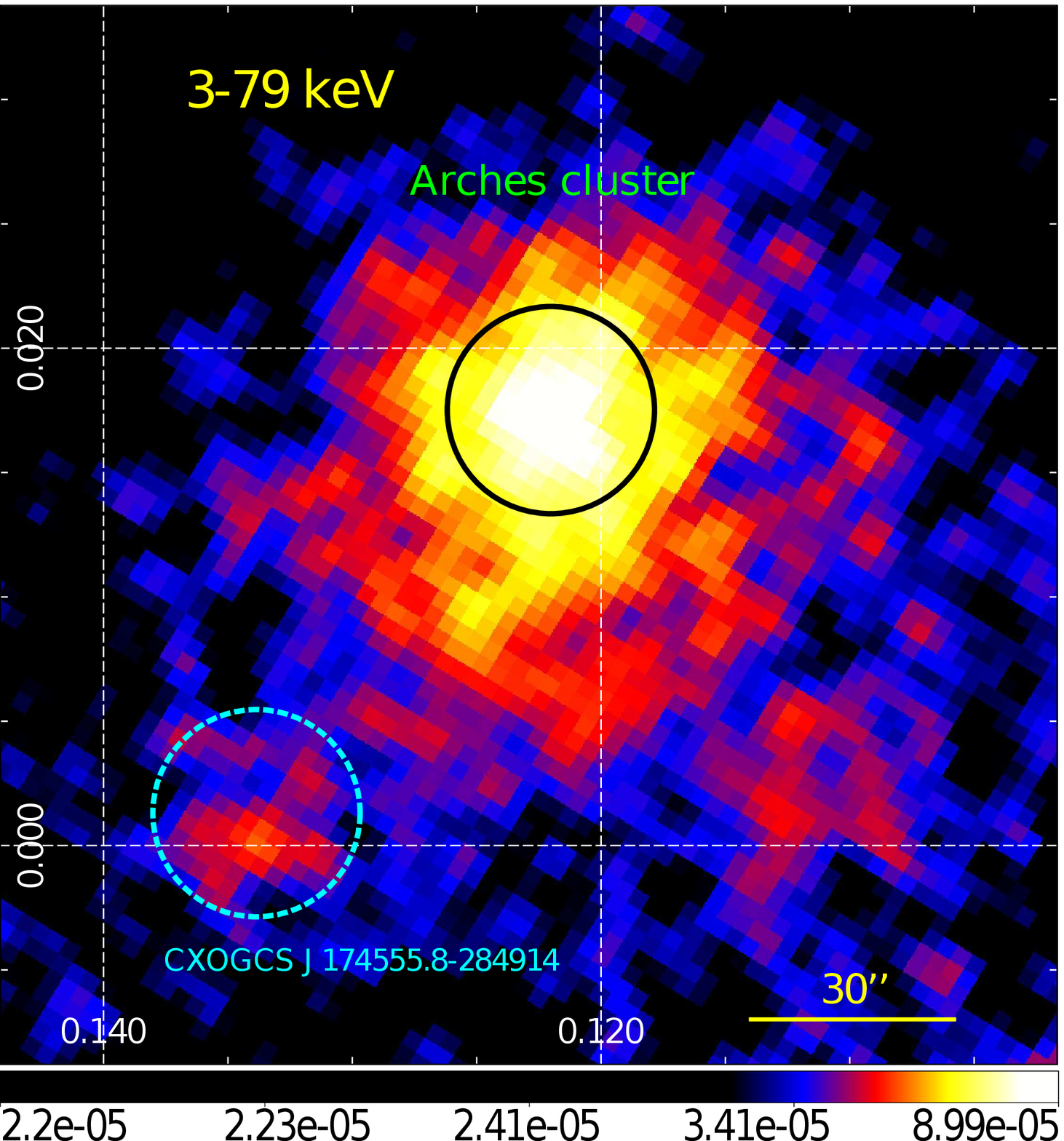}
\includegraphics[width=0.32\textwidth]{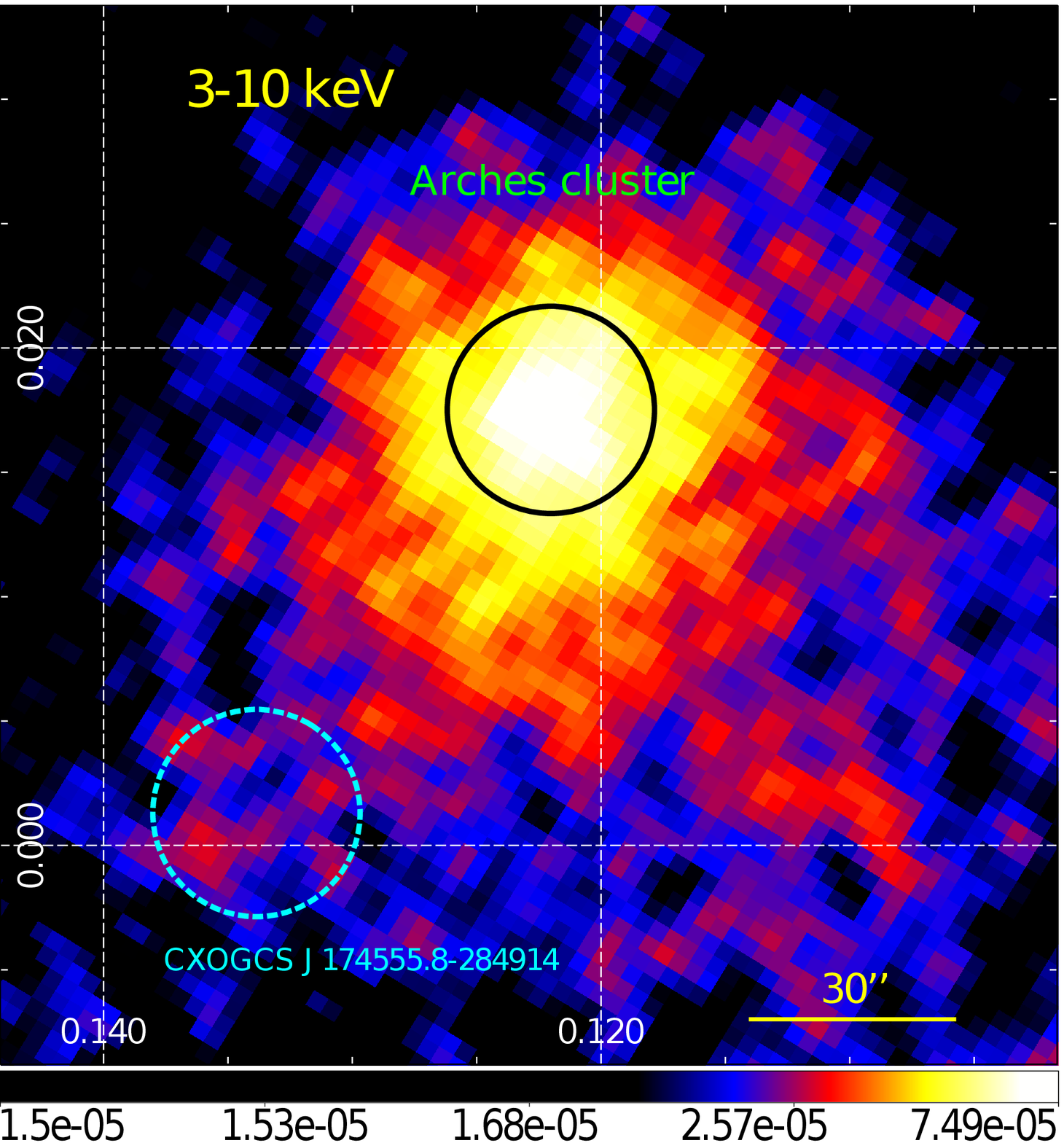} 
\includegraphics[width=0.32\textwidth]{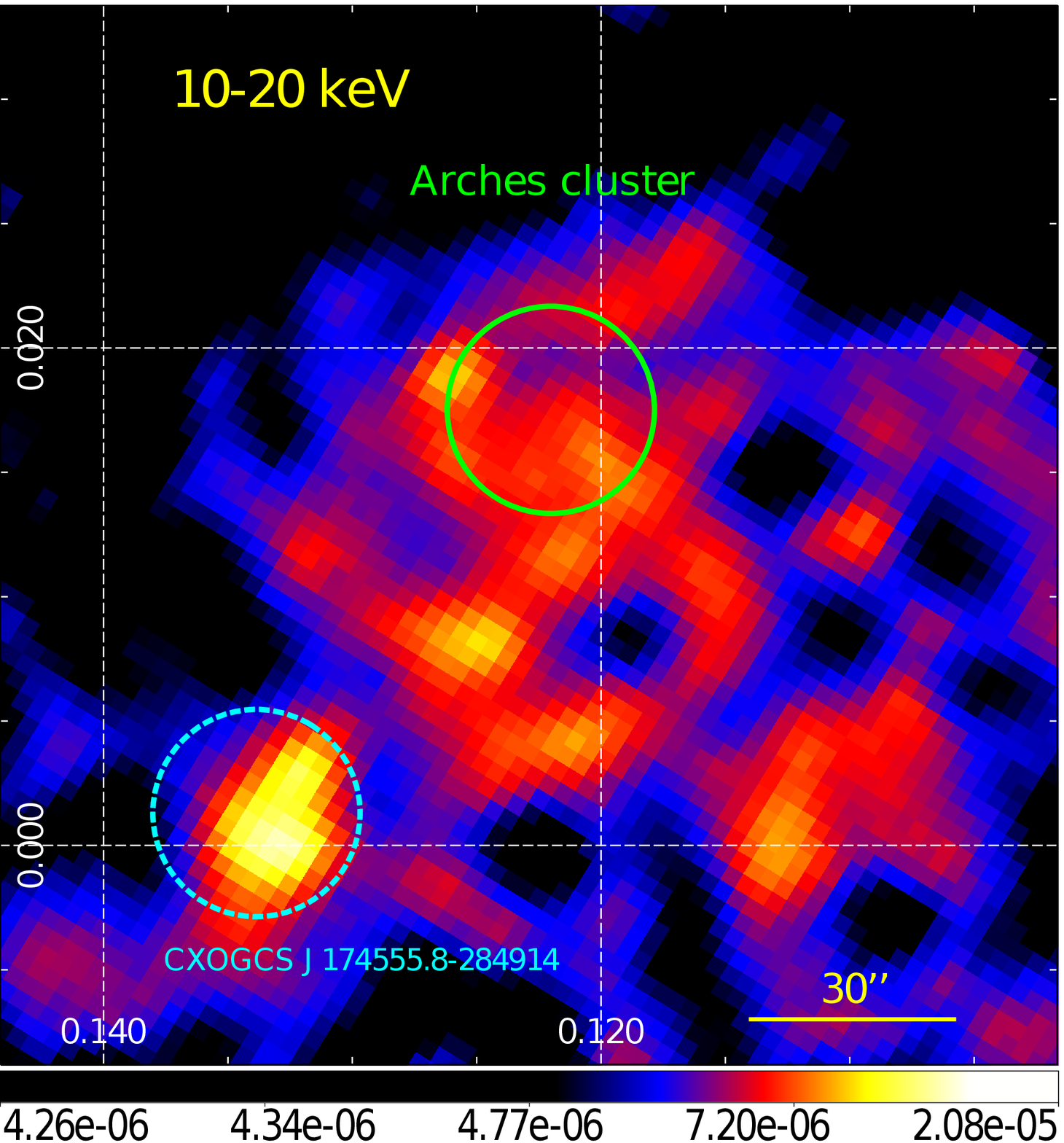} 
\caption{
The \nus\ mosaic images of the Arches cluster region 2015-2016 in $3-79$~keV (left),  $3-10$~keV (center) and $10-20$~keV (right) energy bands. The images are corrected for the exposure and adaptively smoothed with {\it dmimgadapt} task from {\sc ciao-v.4.9} using a tophat kernel with parameters as for Fig.~\ref{fig:fpma}. The sky grid indicates Galactic coordinates in degrees. The solid circle with the radius of $15''$ shows the position of the Arches stellar cluster. The $15''$ dashed circle denotes the position of the hard X-ray point source CXOGCS~J174555.8--284914.}
\label{fig:mosa}
\end{figure*}

Assuming that the morphology of the Arches extended emission has not significantly changed since 2015 (which is confirmed by spectral analysis, see previous section), we combined the \nus\ observations from 2015 and 2016 into one data set, producing mosaic images in three energy bands: $3-79$~keV, $3-10$~keV and $10-20$~keV, shown in Fig.~\ref{fig:mosa}. The full $3-79$~keV band mosaic image reveals both the emission of the stellar cluster and the extended emission. The surface brightness distribution of the Arches in the $3-10$~keV band is dominated by the stellar cluster. Finally, the hard $10-20$~keV band highlights the distribution of the non-thermal emission around the Arches cluster.

\begin{table}
\noindent
\centering
\caption{The 2D-fitting procedure best-fit parameters with the $1\sigma$ errors of the $50''$ central region of the Arches cluster for the $3-79$, $3-10$ and $10-20$~keV energy bands, based on 2015-2016 data.}
\label{tab:rad2d}
\centering
\vspace{1mm}
\begin{tabular}{l r|r|r}
\hline \hline
Parameter & $3-79$~keV & $3-10$~keV & $10-20$~keV \\
\hline 
\multicolumn{4}{c}{Core (2D Gaussian, \texttt{G1})} \\
Center RA (J2000) & 17:45:50.52 & 17:45:50.54 &  \\
Center Dec (J2000) & -28:49:22.39 & -28:49:22.35 &  \\
FWHM & $4''$ (fixed) & $4''$ (fixed) & \\
Norm. ($\times 10^{-3}$) & 3.7$\pm0.5$ & 3.6$\pm0.5$ & \\
\hline
\multicolumn{4}{c}{Cloud (2D Gaussian, \texttt{G2})} \\
Center RA (J2000)  &  17:45:51.47 & 17:45:51.35 & 17:45:52.05 \\
Center Dec (J2000) & -28:49:30.76 & -28:49:29.71 & -28:49:31.72 \\
FWHM & $41''.9\pm3''.5$ & $37''.5\pm3''.5$ & $61''.0^{+14.1}_{-10.0}$ \\
Norm. ($\times 10^{-6}$) & 63.8$\pm11.8$ & 67.0$\pm15.2$ & 6.1$^{+1.7}_{-1.3}$ \\
\hline
\end{tabular}\\
\vspace{3mm}
\end{table}

\begin{figure*}
\center
\includegraphics[width=0.23\textwidth]{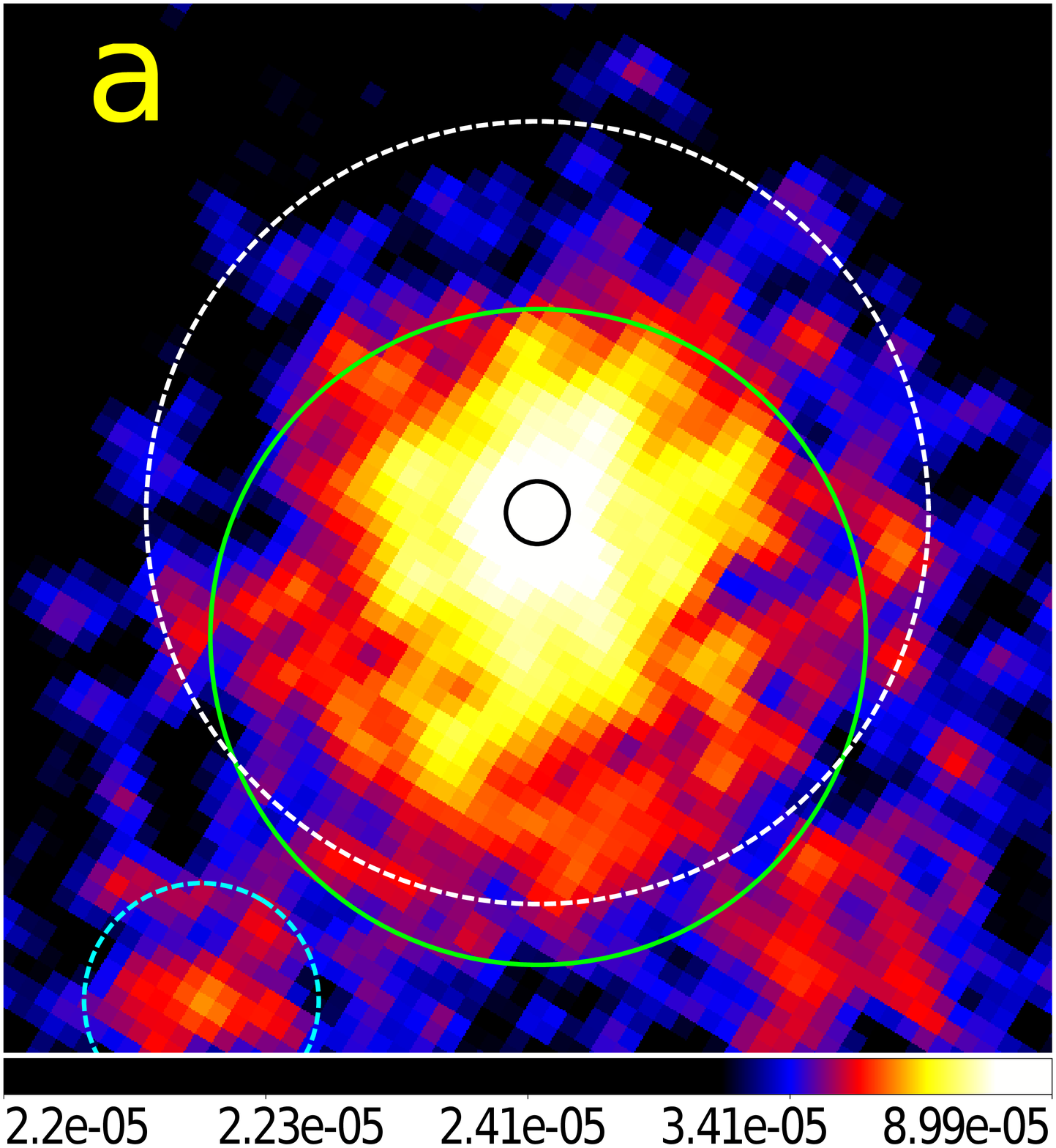}
\includegraphics[width=0.23\textwidth]{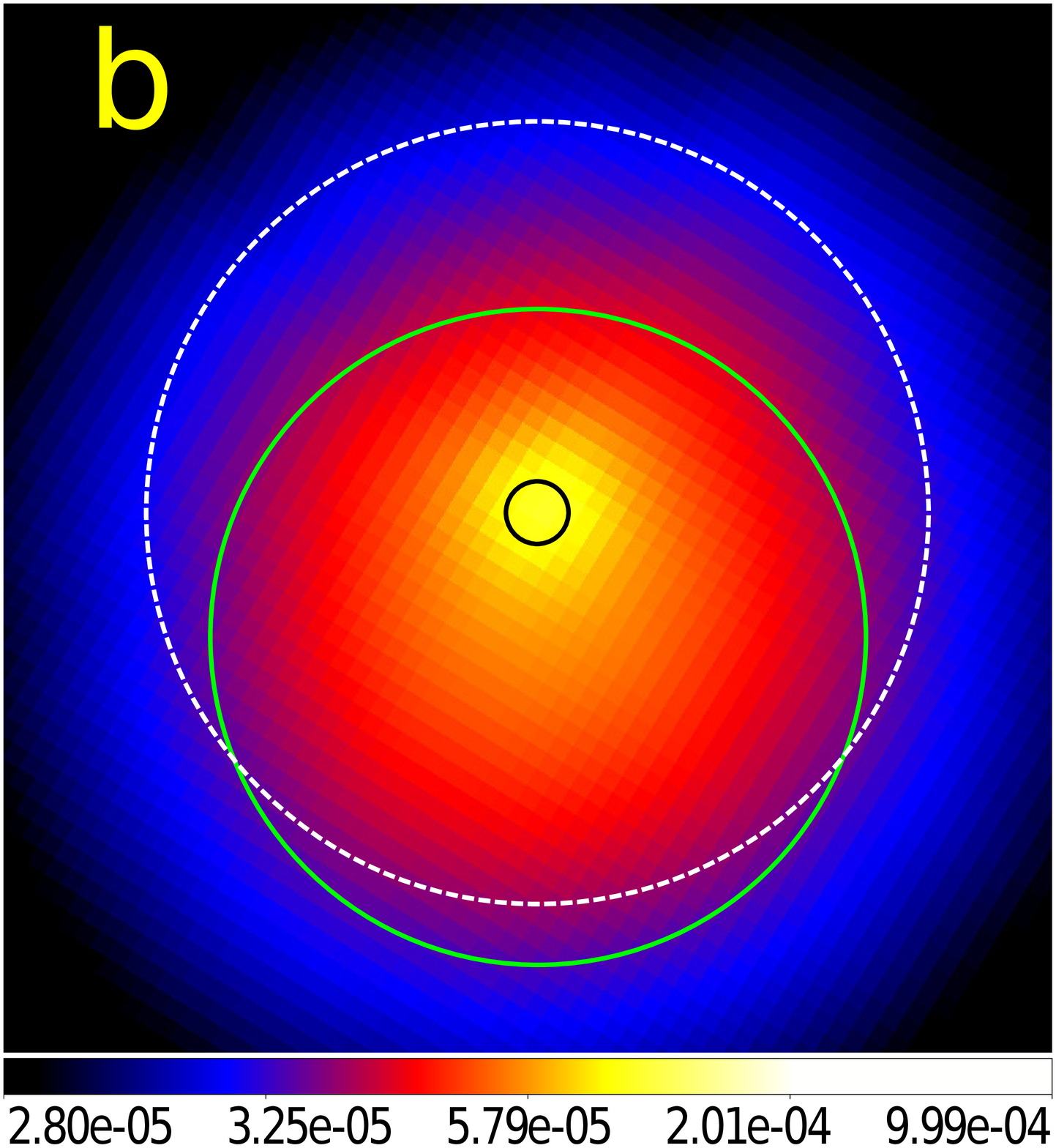}
\includegraphics[width=0.23\textwidth]{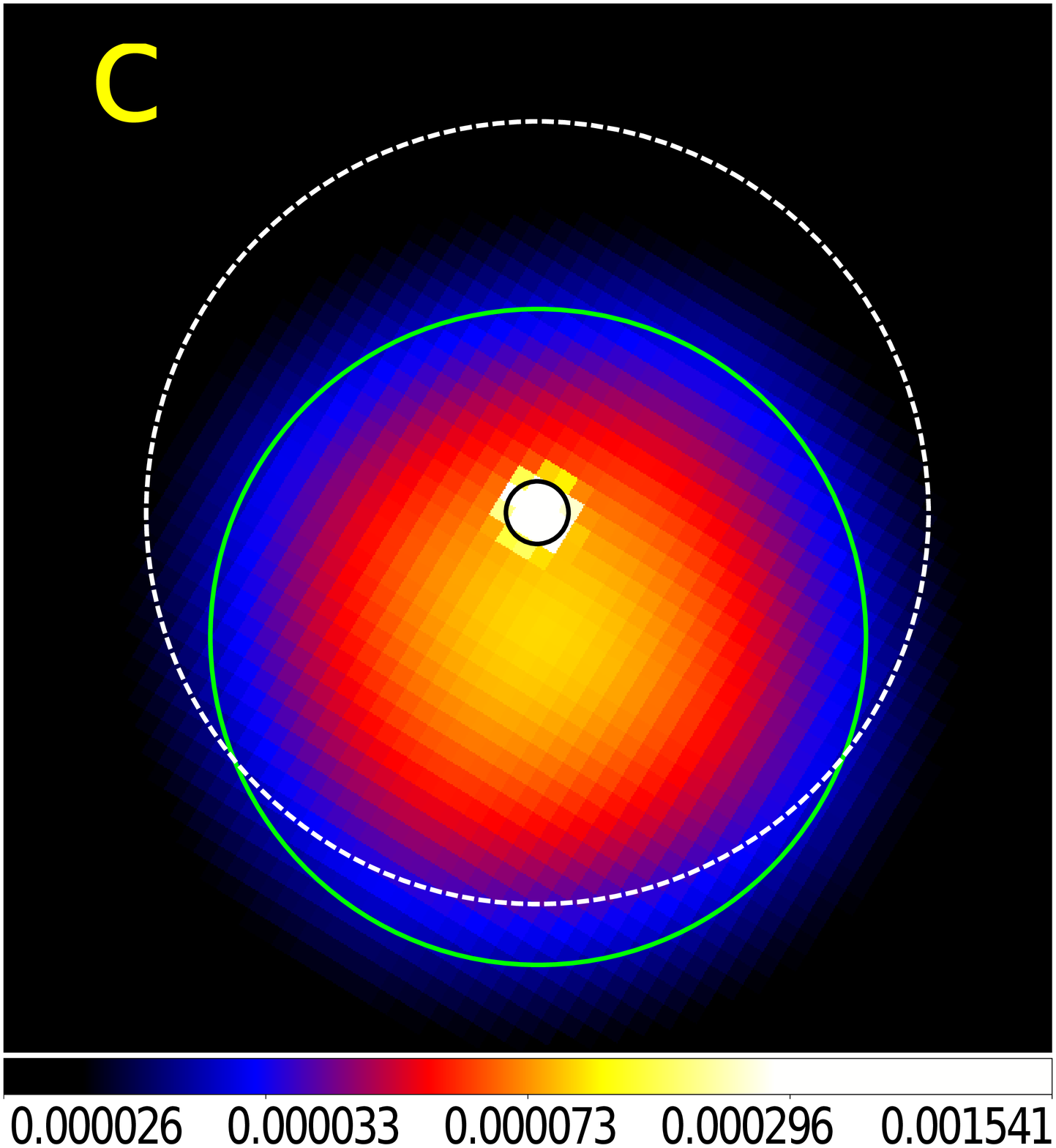}
\includegraphics[width=0.23\textwidth]{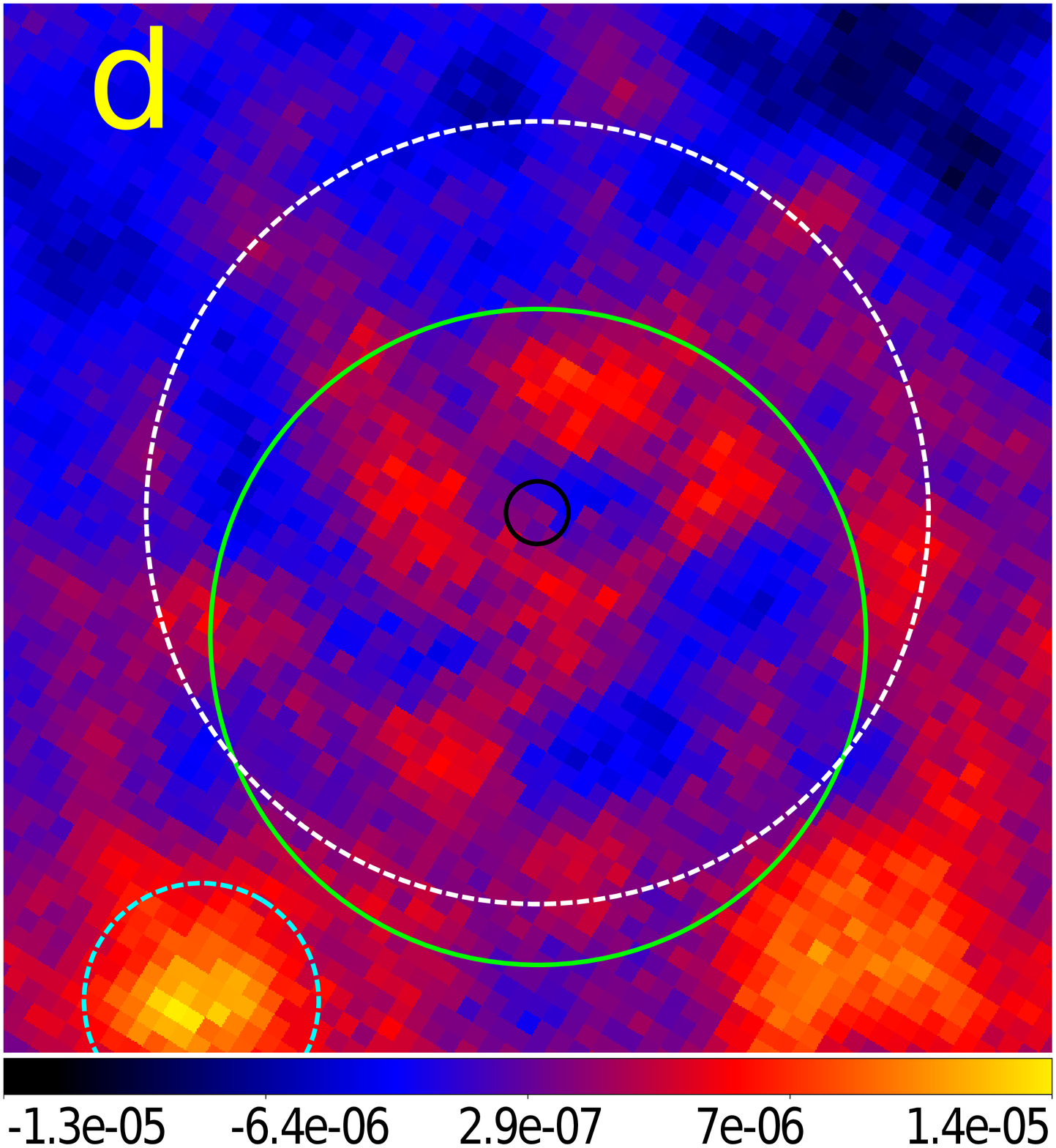}  
\caption{An example of 2D fitting procedure of the Arches cluster mosaic image. Small and large solid regions are \texttt{G1} and \texttt{G2} components of the best-fit Gaussian model with central positions and FWHM radii listed in Table~\ref{tab:rad2d}, respectively. The fitting area constrained by the $R=50''$ circle region (large dashed circle) with excluded $R=15''$ region of the source CXOGCS~J174555.8--284914 (small dashed  circle). a) The $3-79$~keV \nus\ mosaic image. b) The Best-fit spatial model of the Arches cluster complex in $3-79$~keV energy band with two gaussians \texttt{G1} and \texttt{G2} convolved with the NuSTAR PSF. c) The same as plot b, but not convolved with the PSF. d) The residuals after subtracting best-fitting model smoothed with a tophat kernel ($R=3$~px).}
\label{fig:mod2D}
\end{figure*}

We then repeated the 2D image analysis on the combined \nus\ mosaic images, in order to better determine properties of the extended emission. The best-fit model and residuals of the fit for $3-79$~keV energy band image are shown in Fig~\ref{fig:mod2D}. The residual image shows small scale fluctuations with amplitude an order of magnitude less than the input $3-79$~keV \nus\ mosaic image. The positions of the residual fluctuations are spatially consistent with the 6.4~keV Arches cloud clumps described in Sect.~\ref{sec:clumps}. Note that we excluded the \texttt{G1} Gaussian model from the fitting procedure in the $10-20$~keV band due to negligible contribution of the thermal emission from the stellar cluster at energies above 10~keV. The best-fitting model parameters are shown in Table~\ref{tab:rad2d}. In all three bands the spatial size of the \texttt{G2} component is larger than the \nus\ PSF, which confirms the presence of the extended emission around the cluster above 10~keV. We checked that the centroid position of the extended \texttt{G2} component in the $3-10$~keV and $10-20$~keV bands are consistent with each other within $3\sigma$ confidence interval, which points out to the same extended component detected in the soft and the hard energy bands, respectively. 

Note the \texttt{G2} component is more extended in the $10-20$~keV band compared to the $3-10$~keV (see Table~\ref{tab:rad2d}). This is mainly caused by lower statistics of the Arches non-thermal emission above 10~keV and nearby image fluctuations probably associated with the systematic noise. To estimate the level of the systematics we analyzed the source-free region of the $10-20$~keV image and found that the relative non-statistical fluctuations can be as high as $\sim15\%$.

Fig.~\ref{fig:mosa} also reveals an X-ray excess located $\sim1'$ to the east of the cluster, clearly seen in the full $3-79$~keV and hard $10-20$~keV band images. Its centroid position $R.A.=17^h45^m56.0^s$, $Dec.=-28^{\circ}49'16.4''$ is consistent within $3\sigma$ uncertainty with the position of the X-ray source CXOGCS~J174555.8--284914 \citep{zadeh,J174555}, as identified with the SIMBAD astronomical database \citep{simbad}. The source, also labeled as A4 in \cite{zadeh}, is highly absorbed ($N_{\rm H}\sim4\times 10^{23}~cm^{-2}$) and has a hard power-law like spectrum with a peak near the 6.4~keV Fe line \citep{J174555}. To avoid a possible contribution from this source to the background count rate assessment, we excluded $15''$ circular region around it from the spatial and spectral analysis.

\subsection{Spectral analysis}
\label{sec:15-16spec}

\begin{figure}
\center
\includegraphics[width=1\columnwidth]{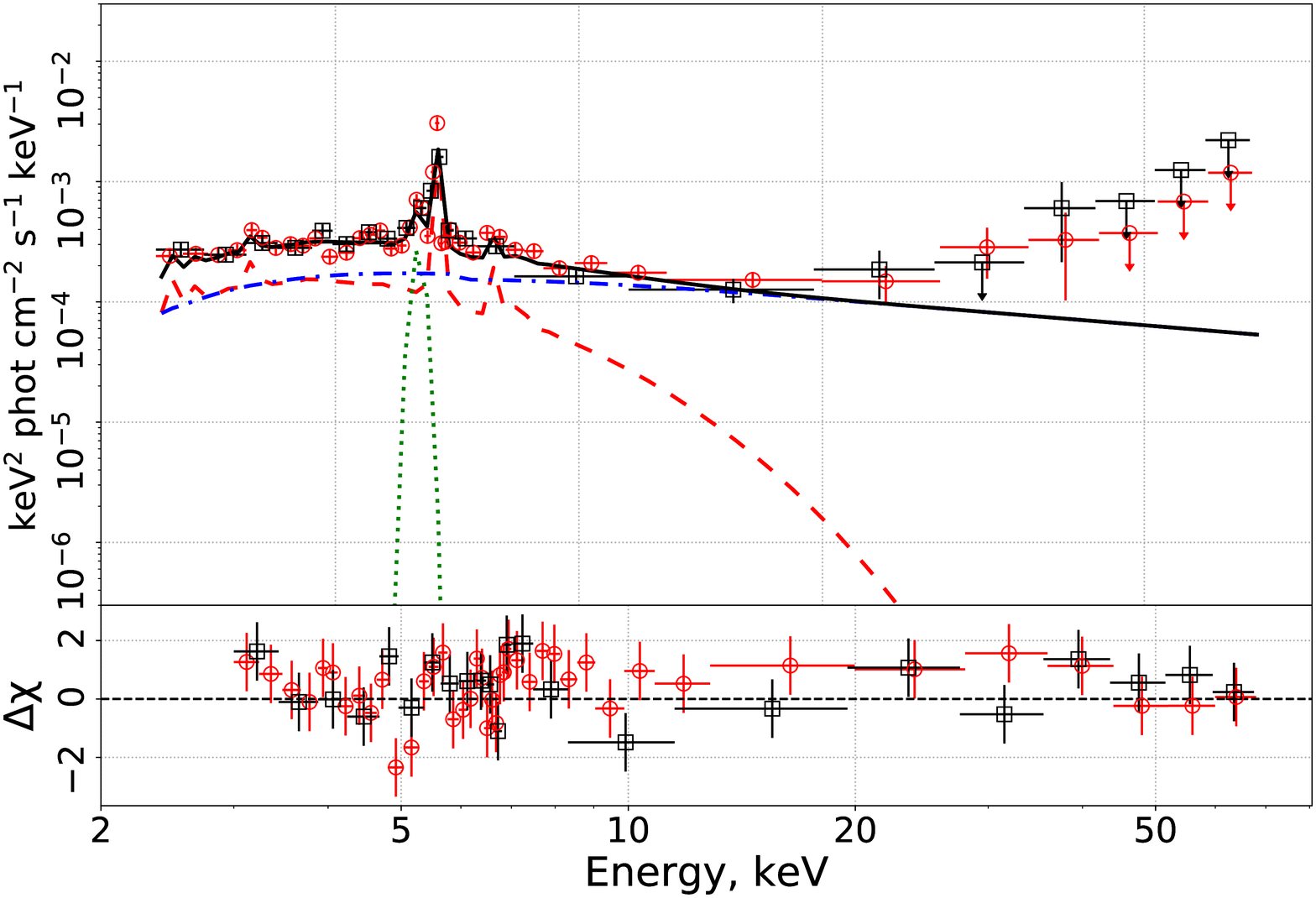}
\caption{Joint 2015-2016 \nus\ X-ray spectrum of the Arches cluster region. 2015 and 2016 \nus\ data are marked with circles and squares, respectively. The total model is represented by a solid line. Dashed, dotted and dash-dotted lines represent thermal plasma, 6.4~keV emission line (width fixed at 0.1~keV) and power-law continuum model components, respectively.}
\label{fig:spec15-16}
\end{figure}

To better estimate the model parameters, we simultaneously fitted the 2015 and 2016 data sets for the following spectral analysis. The last column of Table~\ref{tab:spec} shows the results of the fitting procedure with the same model used in Sect.~\ref{subsec:2016spec}, and Fig.~\ref{fig:spec15-16} demonstrates the combined spectral fit. The best-fitting model parameters of the data sets are in general agreement with individual \nus\ observations in 2015 and 2016.

\begin{figure}
\center
\includegraphics[width=1\columnwidth]{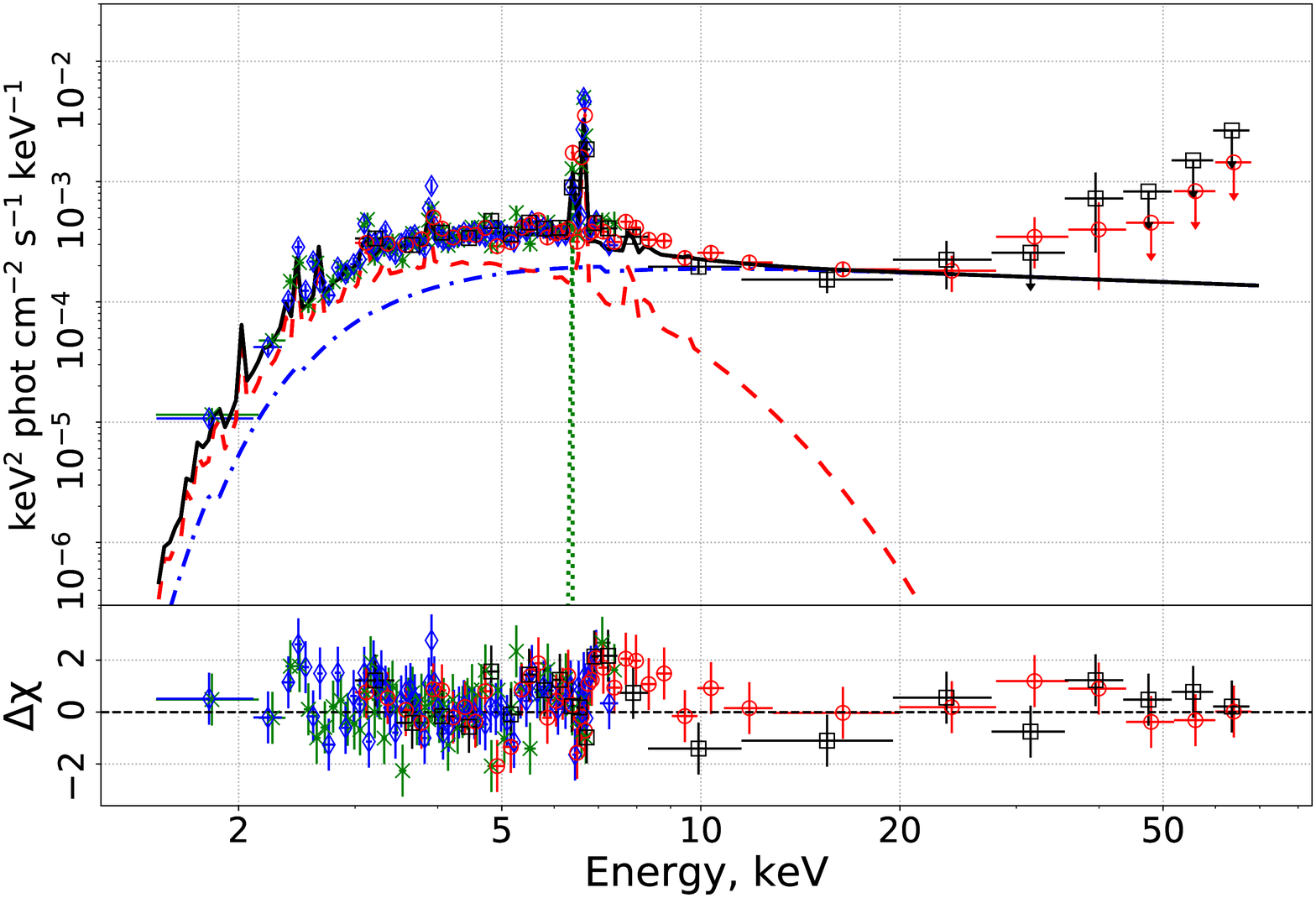}
\caption{Joint \xmm\ 2015 and \nus\ 2015-2016 spectrum of the Arches cluster extracted from the $50''$ circle region around the stellar cluster. Crosses and diamonds represent MOS1/MOS2 and PN data, respectively. Circles and squares show, respectively, 2015 and 2016 \nus/FPMA data. The best-fitting spectral model (solid line, Table~\ref{tab:jointnus_xmm}) includes emission of the thermal plasma (dashed), fluorescent Fe K$\alpha$ line emission (dotted) with a fixed width of 0.01~keV and power-law continuum (dash-dotted).}
\label{fig:specxmm-nus}
\end{figure}

To extend a spectral coverage to the low energies, and to determine the total line-of-sight absorption column density $N_{\rm H}$, we added spectrum based on \xmm\ observation carried out on 27$^{\rm th}$ September 2015 (Table~\ref{tab:obs}), extracted from the same $50''$ circular region \citep{K17}. The combined 2015 \xmm\ and 2015-2016 \nus\ broad-band spectrum of the Arches cluster region is shown in Fig.~\ref{fig:specxmm-nus}. The best-fitting model parameters with the free absorption parameter are listed in Table~\ref{tab:jointnus_xmm}. The $N_{\rm H}$ value was determined as $(9.3\pm0.5)\times10^{22}$~cm$^{-2}$, which is consistent with $N_{\rm H}=(9.5\pm0.3)\times10^{22}$~cm$^{-2}$ measured by \cite{T12}. The joint fit gives a solid measurement of the power-law $\Gamma=2.21\pm0.15$ up to $30-40$~keV, which is in agreement within $1\sigma$ uncertainty with $\Gamma=2.03\pm0.16$ measured by \cite{K17}.

To describe the Arches non-thermal continuum in the combined 2015 \xmm\ and \nus\ spectrum, additionally to a simple power-law model, \cite{K17} also applied CR-induced emission model developed by \cite{T12} and self-consistent X-ray reflection model {\sc REFLIONX} \citep{2005MNRAS.358..211R}. We checked that adding 2016 \nus\ data set to the joint 2015 \xmm\ and \nus\ fit allows to better constrain the parameters of these models, however, the improvements are still not enough to significantly confirm or rule out either of them. For simplicity we do not list the best-fitting results of physically motivated spectral models in this paper.

\begin{table}
\noindent
\centering
\caption{Best-fit parameters for the Arches cluster extracted from $R=50''$ circular region, based on 2015 \xmm\ and 2015-2016 \nus\ observations. For comparison observations are shown best-fit parameters of the 2015 \nus$+${\sc XMM} data \citep{K17}. Model: $wabs\times(APEC + Gaussian + power law)$.}
\label{tab:jointnus_xmm}
\centering
\vspace{1mm}
 \begin{tabular}{c|c|c|c}
\hline\hline
Parameter & Unit & 2015 & 2015$+$2016 \\
\hline
$N_{\rm H}$ & $10^{22}$~cm$^{-2}$ & $9.3^{+0.9}_{-0.5}$ & $9.3\pm0.5$  \\ 
\hline
$kT$ & keV & $1.95\pm0.14$ & $1.95_{-0.12}^{+0.16}$    \\
$I_{\rm kT}$ & see Sect. \ref{subsec:2016spec} & $18\pm3$ & $17\pm3$ \\
\hline
$\Delta E_{\rm 6.4\ keV}$ & keV & 0.01 (fixed) & 0.01 (fixed) \\
$E_{\rm 6.4\ keV}$ & keV & $6.38\pm0.02$ & $6.38\pm0.03$  \\
$F_{\rm 6.4\ keV}$ & $10^{-6}$ ph~cm$^{-2}$~s$^{-1}$ & $3.1\pm0.6$ & $3.0\pm0.6$  \\
$EW_{\rm 6.4\ keV}$ & eV & $700^{+100}_{-90}$ & $660\pm80$ \\
\hline
$\Gamma$&  & $2.03\pm0.16$ & $2.21\pm0.15$ \\
$I_{\rm pow}$ & 10$^{-5}$~cm$^{-2}$~s$^{-1}$~keV$^{-1}$ & $22_{-7}^{+10}$ & $33_{-9}^{+13}$ \\
\hline
C & & $0.82\pm0.04$ & $0.69\pm0.04$ \\
\hline
$\chi^{2}_{\rm r}$/d.o.f.& & 1.00/918 & 0.97/1167 \\
\hline
\end{tabular}\\
\vspace{3mm}
\end{table}

\section{6.4~keV clumps of the Arches cloud}
\label{sec:clumps}

The broad-band \nus\ spectral analysis of the Arches non-thermal X-ray emission has been previously performed in a wide circular region covering the stellar cluster itself and surrounding molecular cloud \citep[see Sections \ref{sec:2016} and \ref{sec:15-16},][and]{K14,K17}. Other studies, mainly based on \xmm\ observations, used an elliptical region excluding the stellar cluster \citep[see ellipse parameters in][]{T12,clavel14,K17}. Thus, the spectral extraction region often included the whole Arches non-thermal emission, which was characterized by regular and extended shape, traced by iron 6.4~keV line flux \citep{T12}. Note that integrating the signal over large region may average emission having different trend and$/$or different origin \citep[see e.g.][]{clavel13}. The smaller ellipse which was used in the previous \xmm\ analysis may still be too large to understand the complexity of the current $EW$ variation. To investigate this we carried out new analysis of the clumps revealed in the 6.4~keV line using the 2015 \xmm\ data.

\begin{table*}
\noindent
\centering
\caption{Definitions of the sky regions of the Arches clumps used for the spectral analysis. }
\label{tab:clumps_reg}
\centering
\vspace{1mm}
\begin{tabular}{c|c|c|c|c}
\hline
\hline
Region & RA (J2000) & Dec. (J2000) & Parameters & Area, arcsec$^2$ \\
\hline
Clump S & 17:45:52.92 & -28:50:21.30 & 26$''$, 44$''$, 130$^{\circ}$ & 3556 \\ 
Clump N & 17:45:49.24 & -28:48:54.65 & 20$''$, 33$''$, 169$^{\circ}$ & 1437$^1$ \\
Clump E & 17:45:52.82 & -28:49:25.35 & 13$''$, 22$''$, 162$^{\circ}$ & 906 \\
Excluded cluster & 17:45:50.47 & -28:49:15.70 & 16$''$, 29$''$, 177$^{\circ}$ & \\
\hline
$^1$) The surface area of clump N is computed with the cluster region excluded.
\end{tabular}
\vspace{3mm}
\end{table*}

\begin{figure*}
\center
\includegraphics[width=1\columnwidth]{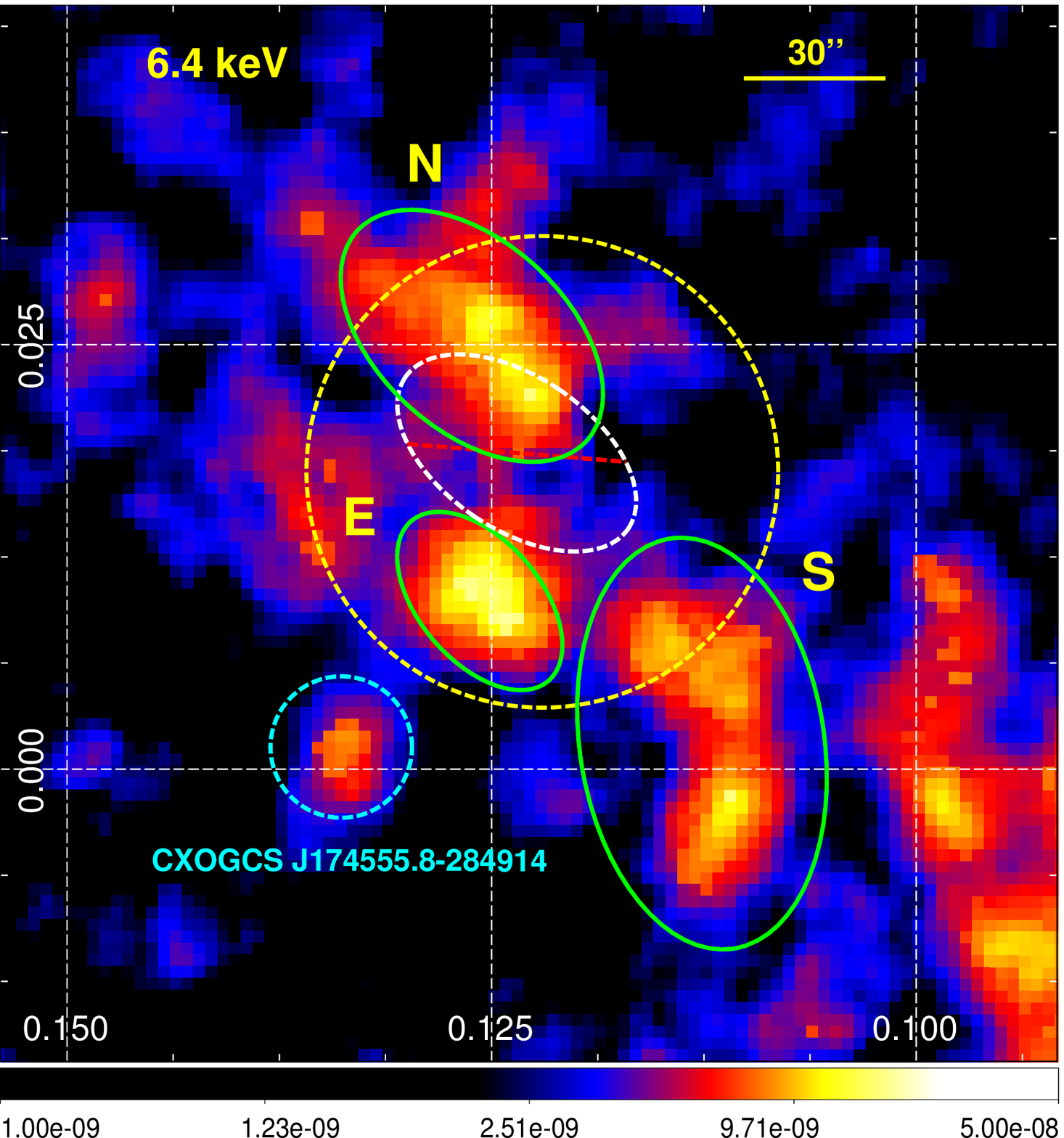} 
\includegraphics[width=1\columnwidth]{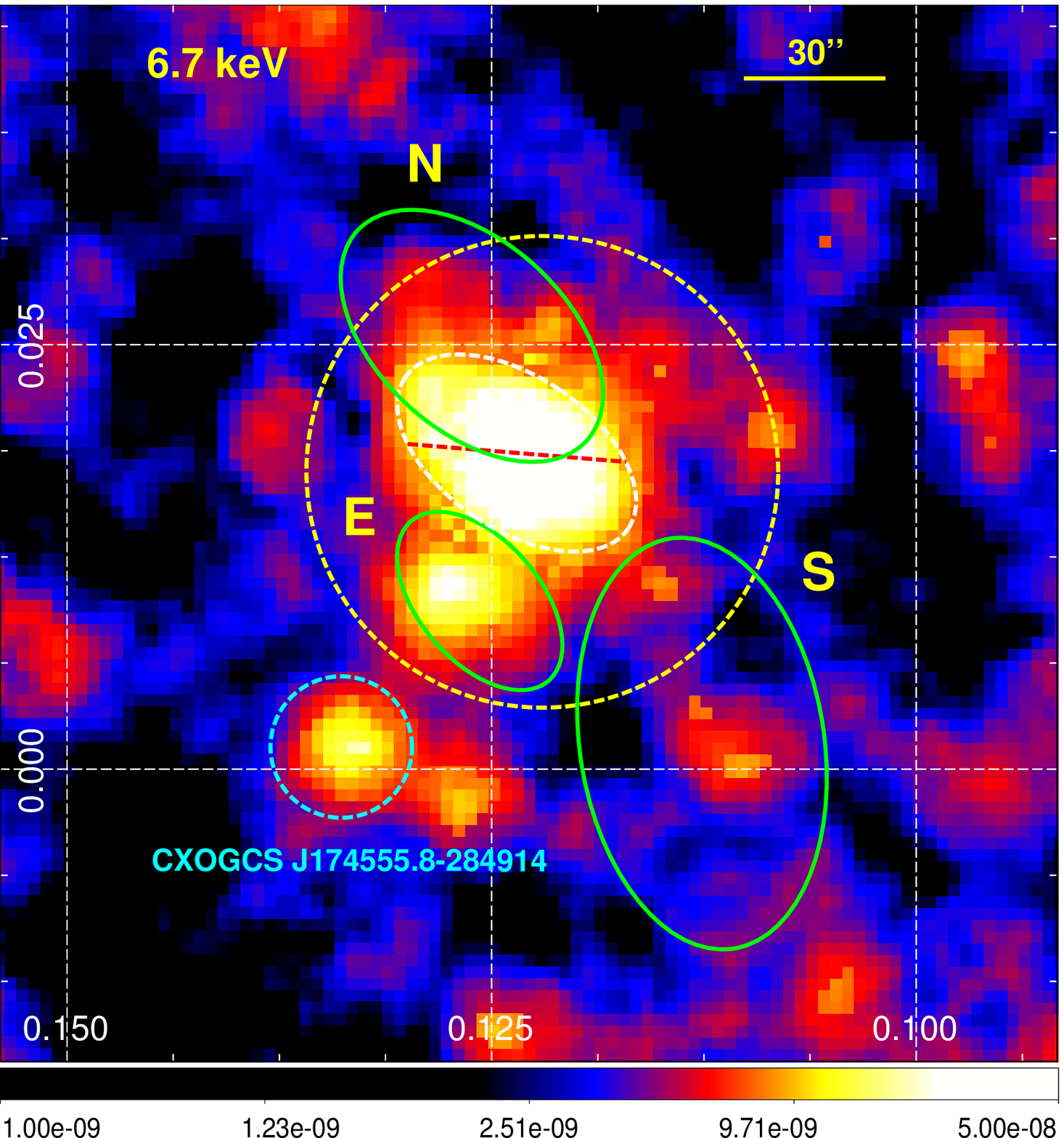} 
\caption{The 6.4~keV (left) and 6.7~keV (right) line flux mosaic images constructed, respectively, in the $6.32-6.48$ and $6.62-6.78$~keV energy bands \citep[\xmm, 2015,][]{K17}. The maps are continuum subtracted and shown in Galactic coordinates. The solid ellipses represent the source spectrum extraction regions of the Arches cloud clumps. The cluster core emission bright in 6.7~keV line is shown by the dashed ellipse. The small dashed circle ($R=15''$) indicates the position of the hard X-ray source CXOGCS~J174555.8--284914 \citep{zadeh,J174555}. The large dashed circle region ($R=50''$) shows the source spectrum extraction region of the Arches cluster non-thermal emission (see Sect.~\ref{sec:obs}). The images are adaptively smoothed using the {\sc dmimgadapt} task from {\sc ciao-v.4.9} with the following parameters: tophat kernel, smoothing scales $0.1-5$, the number of scales is 10, the minimum number of counts is 20.}
\label{fig:clumps_reg}
\end{figure*}

\begin{figure}
\center
\includegraphics[width=1\columnwidth]{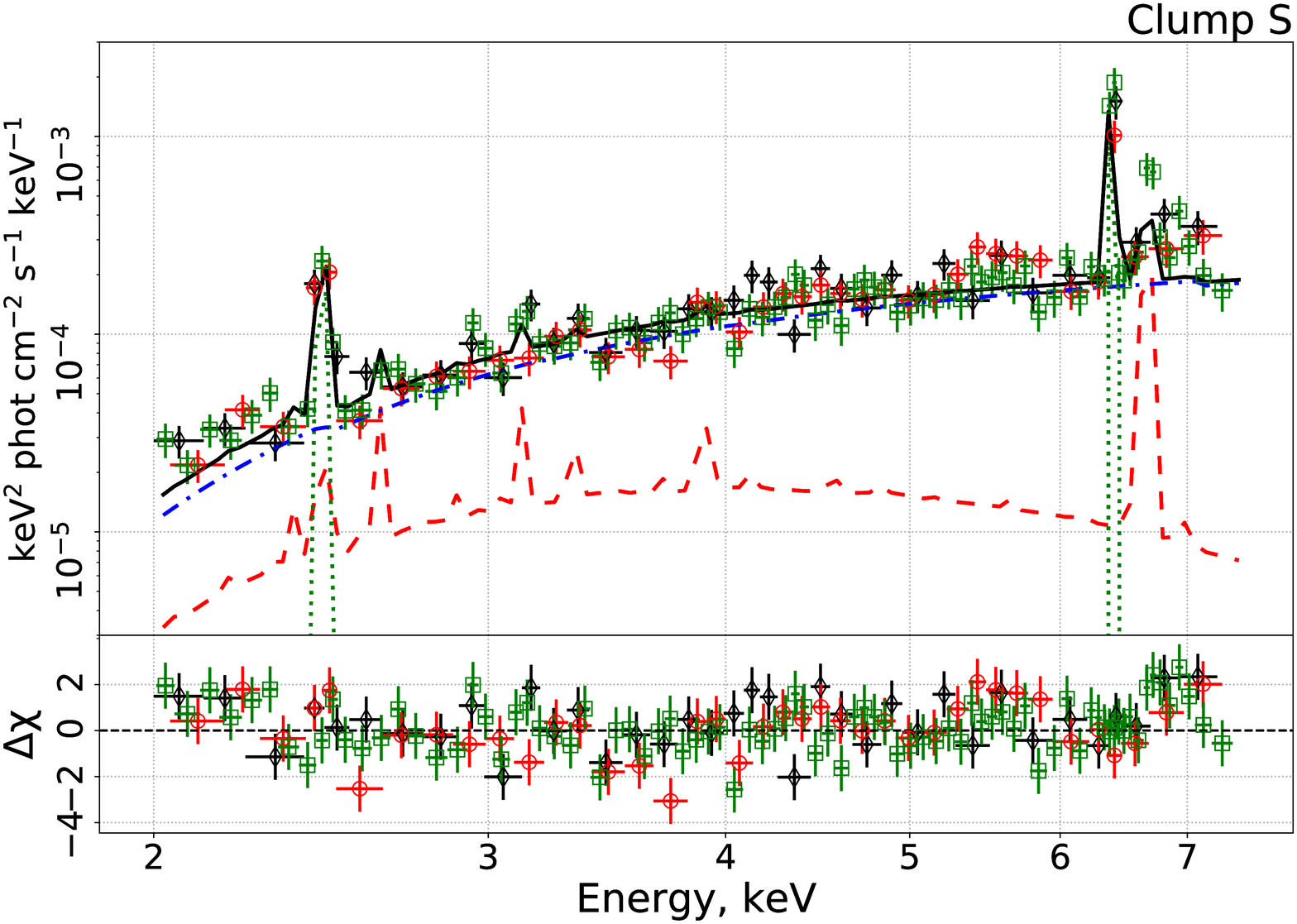} 
\includegraphics[width=1\columnwidth]{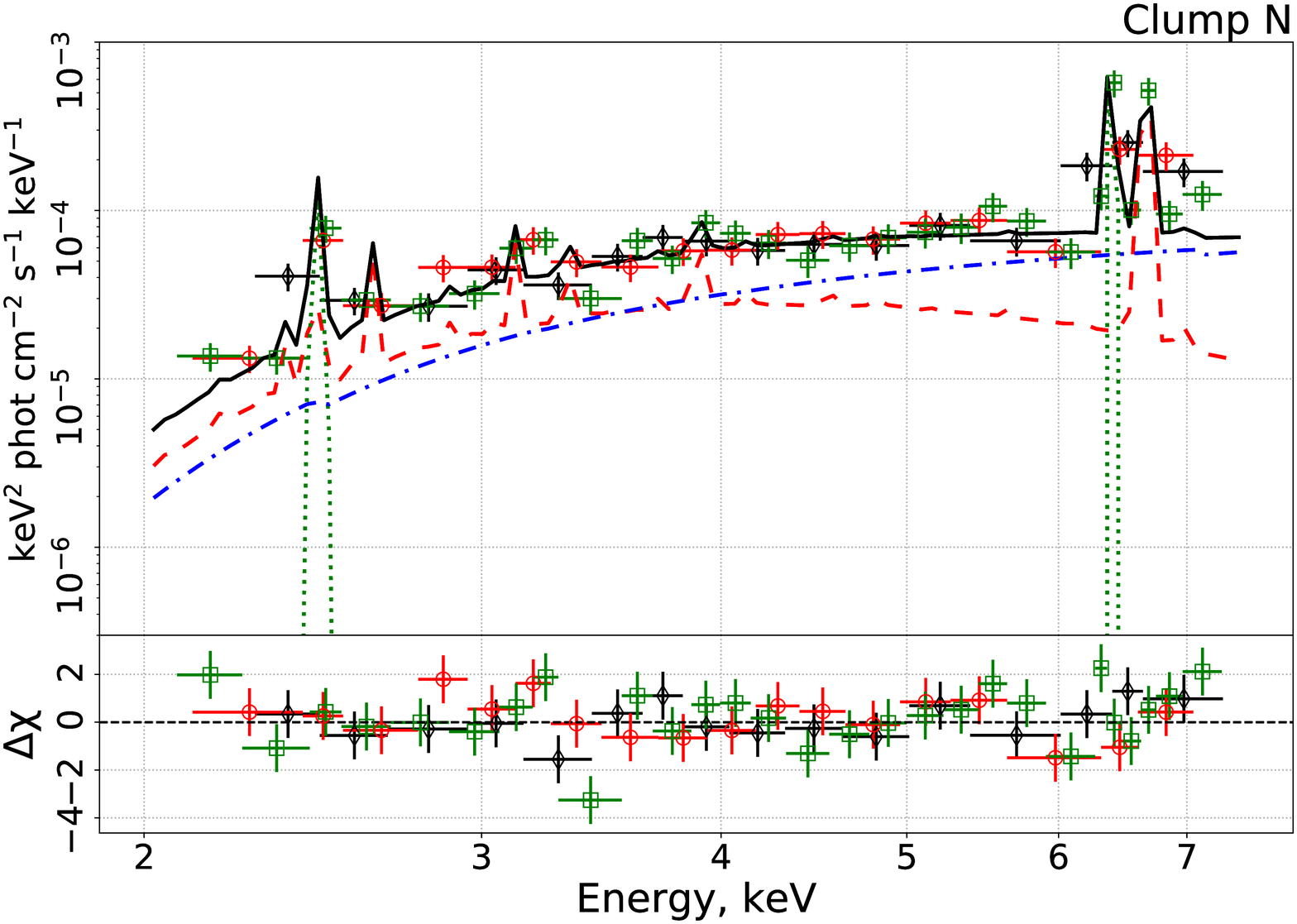} 
\includegraphics[width=1\columnwidth]{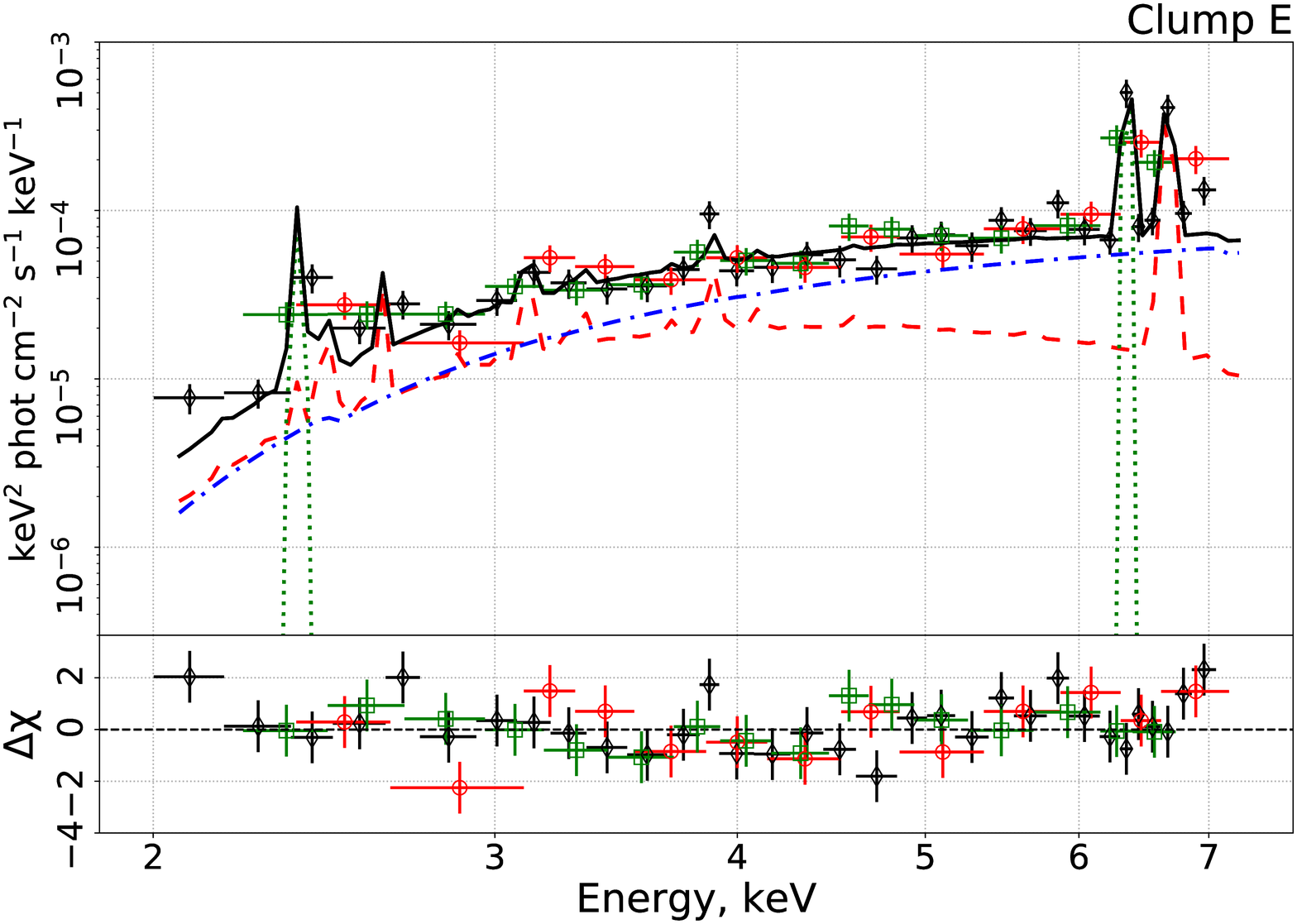} 
\caption{The spectra of each clump obtained with \xmm\ in 2015. Diamonds, circles and squares represent MOS1, MOS2 and PN data, respectively. The spectra were extracted in the $2-7.5$~keV energy range for all clumps except clump E for which energy range was $2-7.3$~keV due to low statistics. The solid line represents best-fit model ($wabs \times (APEC + Gaussian^{\rm 2.45\ keV} + power law) + Gaussian^{\rm 6.4\ keV}$) with the following components: thermal plasma (dashed), Sulfur 2.45~keV and Fe 6.4~keV line emission (dotted) and non-thermal power-law continuum (dash-dotted).}
\label{fig:clumps}
\end{figure}

We followed the procedure described in \cite{K17} for analyzing the \xmm\ MOS1, MOS2 and PN data and extracted $2-7.5$~keV spectra of three clumps S, N and E (see Fig.~\ref{fig:clumps_reg})  within elliptical regions, which parameters are listed in Table~\ref{tab:clumps_reg}. The regions are shown in Fig.~\ref{fig:clumps_reg}. Note that for the clump N region we excluded the elliptical region containing the bright 6.7~keV line emission from the Arches cluster core. The spectra of clumps are shown in Fig.~\ref{fig:clumps}. Each spectrum was approximated with the {\sc xspec} model $wabs \times (APEC + Gaussian^{\rm 2.45\ keV} + power law) + Gaussian^{\rm 6.4\ keV}$,  consistent with previous \xmm\ study of the Arches non-thermal emission by \cite{clavel14}. The spectral model includes the APEC thermal emission of the astrophysical plasma with $kT=2.2$~keV \citep[chosen to mimic the combination of 1 and 7~keV plasma components in the Galactic center,][]{clavel14}, non-thermal power-law continuum with a fixed photon index $\Gamma=1.6$ and unabsorbed Fe \ka\ 6.4~keV line with a fixed width of 0.01~keV. Following \cite{K17} we also added an absorbed Gaussian line at 2.45~keV, presumably belonging to the \ka\ line from He-like Sulfur (S). The strength of photo-electric absorption is freely fitted for each clump individually. The equivalent width of the Gaussian lines was calculated with respect to the power-law component without absorption. The results of the fitting procedure are shown in Table~\ref{tab:clumps}.

The spectral properties of N and E emission are similar, including $EW_{\rm 6.4\ keV}$, which indicates the same origin of the cloud emission. The clump S has the highest flux in the 6.4~keV line and continuum, however its $EW_{\rm 6.4\ keV}$ value ($570\pm50$~eV) is smaller than for N and E ($980\pm170$~eV and $830\pm200$, respectively). From the other side the clump S has the largest size compared to the others. To estimate the surface brightness we divided the observed $2-7.5$~keV flux of each clump by the corresponding source extraction area as listed in Table~\ref{tab:clumps}.

\begin{table*}
\noindent 
\begin{center}
\caption{Best-fitting model parameters of three non-thermal emission clumps around the Arches stellar cluster measured with \xmm\ in 2015. The model is described in {\sc xspec} notation as $wabs \times (APEC + Gaussian^{\rm 2.45\ keV} + power law) + Gaussian^{\rm 6.4\ keV}$.}
\label{tab:clumps}
\vspace{1mm}
\begin{tabular}{c|c|c|c|c}
\hline \hline
Parameter$^{1)}$ & Unit & Clump S & Clump N & Clump E \\
\hline
$N_{\rm H}$  & $10^{22}$~cm$^{-2}$      & 5.5$\pm0.5$  & 7.2$\pm1.0$ & 8.2$\pm1.1$ \\
\hline
$I_{\rm kT}$ & see Sect. \ref{subsec:2016spec} & 0.9$\pm0.5$ & 1.7$\pm0.6$ &  1.3$\pm0.5$\\
\hline
$E_{\rm 2.45\ keV} $ & keV                             & 2.45$\pm0.01$ & 2.46$\pm0.03$ & 2.37$_{-0.05}^{+0.07}$ \\
$N_{\rm 2.45\ keV} $ & $10^{-6}$ ph~cm$^{-2}$~s$^{-1}$ & 5.3$_{-1.5}^{+1.8}$ & 4.3$_{-2.1}^{+3.0}$ & 4.9$_{-2.8}^{+4.0}$ \\
$EW_{\rm 2.45\ keV}$ & eV                              & 260$\pm40$ & 710$\pm150$ & 1500$\pm400$\\
\hline
$E_{\rm 6.4\ keV}  $ & keV                             & 6.39$\pm0.02$ & 6.39$\pm0.03$ & 6.35$\pm0.03$ \\
$N_{\rm 6.4\ keV}  $ & $10^{-6}$ ph~cm$^{-2}$~s$^{-1}$ & 2.6$\pm0.4$ & 1.4$\pm0.3$ & 1.2$\pm0.3$ \\
$EW_{\rm 6.4\ keV} $ &  eV                             & 570$\pm50$ & 980$\pm170$ & 830$\pm200$\\
\hline
$N^{\rm pow}_{\rm @keV}$          & $10^{-5}$ ph~keV$^{-1}$~cm$^{-2}$~s$^{-1}$ & 9.1$\pm0.8$ & 2.9$\pm0.7$ & 3.0$\pm0.6$ \\
\\[-1em]
$F^{\rm pow}_{\rm 2-7.5\ keV}$      & $10^{-13}$~erg~cm$^{-2}$~s$^{-1}$ & 3.35$\pm0.28$ & 1.08$^{+0.25}_{-0.27}$ & 1.30$\pm0.22$ \\
\\[-1em]
$S^{\rm pow}_{\rm 2-7.5\ keV}$        & $10^{-17}$~erg~cm$^{-2}$~s$^{-1}$~arcsec$^{-2}$ & 9.4$\pm0.8$ & 7.5$^{+1.7}_{1.9}$ & 14.3$\pm2.4$ \\
\hline
C & & 1.02$\pm0.05$ & 0.72$\pm0.06$ & 1.1$\pm0.1$ \\
$\chi^2/d.o.f.$ & & 1.33$/$151 & 1.18$/$54 & 1.07$/$53\\
\hline
\end{tabular}\\
\vspace{3mm}
\end{center}
$^{1)}$ S$^{\rm pow}_{\rm 2-7.5\ keV}$ is the flux in the $2-7.5$~keV band divided on the clump region area. C denotes cross-normalization factor between MOS1$/$MOS2 and PN data. \\
\end{table*}

\section{Discussion and summary}
\label{sec:summ}

In this paper we present results of a long \nus\ (150 ks) observation of the Arches complex in 2016. The complex includes a stellar cluster with a bright thermal ($kT\sim2$~keV) X-ray emission and a nearby molecular cloud, characterized by an extended non-thermal X-ray continuum and fluorescent Fe \ka\ 6.4~keV line. Recent \nus\ and \xmm\ observations of the Arches non-thermal emission demonstrate a dramatic change both in morphology and intensity, and continuous observations are needed to shed light on the nature of the current low-level emission.

Despite the fast decrease observed in 2007-2015 \citep{K17}, the \nus\ observations in 2016 show that the non-thermal emission is still present around the Arches stellar cluster, as demonstrated by our spectral and spatial analysis. Furthermore, the $3-20$~keV flux of the emission in 2016 remains at the same level as measured in 2015 without a significant change in the spectral shape. However, one year interval may not be long enough to see a strong flux variation, which is followed from the fact that the measured 2016 non-thermal flux is not in contradiction with the declining trend observed in 2007-2015 \citep{K17}. The overall drop of the $EW_{\rm 6.4\ keV}$ from $\sim0.9$~keV in 2002-2015 to $0.6-0.7$~keV in 2015 \citep{K17} has been confirmed at the same level in 2016. These results may indicate that the intensity of the Arches non-thermal emission reached its stationary level. At the same time, the declining trend predicts the 2016 flux at the level of $F_{\rm 3-20\ keV}=5.2^{+1.1}_{-0.9}\times 10^{-13}$~erg~cm$^{-2}$~s$^{-1}$, which is consistent within the errors with our measurement $F_{\rm 3-20\ keV}=4.4^{+2.1}_{-3.1}\times 10^{-13}$~erg~cm$^{-2}$~s$^{-1}$ in 2016 (Table~\ref{tab:spec}). Thus, we cannot assess whether the non-thermal emission has reached a stationary level in 2016, and new observations, separated by a longer time period, are needed.

To better constrain the spectral properties of the Arches complex in this state, we constructed its broad-band spectrum using \nus\ 2015-2016 and \xmm\ 2015 observations, with the total exposure time of 350~ks and 114~ks, respectively. The joint fit gives a solid measurement of the power-law $\Gamma=2.21\pm0.15$ up to $30-40$~keV, which is, however softer than $\Gamma=2.03\pm0.16$ measured by \cite{K17} in 2015 with the same \xmm\ data set.

The rough 2D spatial analysis with \nus\ shows that the morphology of the non-thermal emission in 2016 is still associated with molecular cloud and is not consistent with the possible emission of the stellar cluster.

Finally, we performed the spectral analysis of three bright Fe $K_{\alpha}$ line emission clumps of the Arches molecular cloud \citep{K17}, based on \xmm\ observations in 2015. The analysis showed different $EW_{\rm 6.4\ keV}$ values for clumps S ($570\pm50$~eV) and N-E ($\sim900$~eV). The clump S also demonstrates a weaker interstellar absorption than two others. This finding confirms that the X-ray emission from the molecular cloud is a combination of two components with different origins. The different $EW$s of the iron~\ka\ line may also be produced by the different reflection geometry of the clumps. \cite{2018ApJ...863...85C} favor the scenario of two \sgra\ flaring events, consistent with different 6.4~keV line $EW$s found in this work. This is also in agreement with the studies of other molecular clouds in the CMZ, which imply two \sgra\ flares \citep{clavel13,2018A&A...612A.102T,chuard18}.

Further monitoring observations of the Arches complex are strongly needed to trace the on-going evolution of the clumps and to confirm the reached stationary level of the non-thermal emission of the whole Arches cloud region.

\section*{Acknowledgments}

This work has made use of data from the \nus\ mission, a project led by the California Institute of Technology, managed by the Jet Propulsion Laboratory and funded by NASA, and observations obtained with \xmm, an ESA science mission with instruments and contributions directly funded by ESA Member States and NASA. The research has made use of the \nus\ Data Analysis Software jointly developed by the ASI Science Data Center (ASDC, Italy) and the California Institute of Technology. EK and RK acknowledge support from the Russian Basic Research Foundation (grant 16-02-00294). GP acknowledges financial support from the Bundesministerium f\"{u}r Wirtschaft und Technologie/Deutsches Zentrum f\"{u}r Luft- und Raumfahrt (BMWI/DLR, FKZ 50 OR 1812, OR 1715 and OR 1604) and the Max Planck Society. DC is supported by the grant RFBR 18-02-00075.

\bibliographystyle{mnras}
\bibliography{references}

\begin{thebibliography}{}
\makeatletter
\relax
\def\mn@urlcharsother{\let\do\@makeother \do\$\do\&\do\#\do\^\do\_\do\%\do\~}
\def\mn@doi{\begingroup\mn@urlcharsother \@ifnextchar [ {\mn@doi@}
  {\mn@doi@[]}}
\def\mn@doi@[#1]#2{\def\@tempa{#1}\ifx\@tempa\@empty \href
  {http://dx.doi.org/#2} {doi:#2}\else \href {http://dx.doi.org/#2} {#1}\fi
  \endgroup}
\def\mn@eprint#1#2{\mn@eprint@#1:#2::\@nil}
\def\mn@eprint@arXiv#1{\href {http://arxiv.org/abs/#1} {{\tt arXiv:#1}}}
\def\mn@eprint@dblp#1{\href {http://dblp.uni-trier.de/rec/bibtex/#1.xml}
  {dblp:#1}}
\def\mn@eprint@#1:#2:#3:#4\@nil{\def\@tempa {#1}\def\@tempb {#2}\def\@tempc
  {#3}\ifx \@tempc \@empty \let \@tempc \@tempb \let \@tempb \@tempa \fi \ifx
  \@tempb \@empty \def\@tempb {arXiv}\fi \@ifundefined
  {mn@eprint@\@tempb}{\@tempb:\@tempc}{\expandafter \expandafter \csname
  mn@eprint@\@tempb\endcsname \expandafter{\@tempc}}}

\bibitem[\protect\citeauthoryear{{Arnaud}}{{Arnaud}}{1996}]{xspec}
{Arnaud} K.~A.,  1996, in {Jacoby} G.~H.,  {Barnes} J.,  eds,  Astronomical
  Society of the Pacific Conference Series Vol. 101, Astronomical Data Analysis
  Software and Systems V. p.~17

\bibitem[\protect\citeauthoryear{{Capelli}, {Warwick}, {Cappelluti},
  {Gillessen}, {Predehl}, {Porquet}  \& {Czesla}}{{Capelli}
  et~al.}{2011a}]{capellia}
{Capelli} R.,  {Warwick} R.~S.,  {Cappelluti} N.,  {Gillessen} S.,  {Predehl}
  P.,  {Porquet} D.,   {Czesla} S.,  2011a, \mn@doi [\aap]
  {10.1051/0004-6361/201015758}, \href
  {http://adsabs.harvard.edu/abs/2011A%26A...525L...2C} {525, L2}

\bibitem[\protect\citeauthoryear{{Capelli}, {Warwick}, {Porquet}, {Gillessen}
  \& {Predehl}}{{Capelli} et~al.}{2011b}]{capellib}
{Capelli} R.,  {Warwick} R.~S.,  {Porquet} D.,  {Gillessen} S.,   {Predehl} P.,
   2011b, \mn@doi [\aap] {10.1051/0004-6361/201116574}, \href
  {http://adsabs.harvard.edu/abs/2011A%26A...530A..38C} {530, A38}

\bibitem[\protect\citeauthoryear{{Chernyshov}, {Ko}, {Krivonos}, {Dogiel}  \&
  {Cheng}}{{Chernyshov} et~al.}{2018}]{2018ApJ...863...85C}
{Chernyshov} D.~O.,  {Ko} C.~M.,  {Krivonos} R.~A.,  {Dogiel} V.~A.,   {Cheng}
  K.~S.,  2018, \mn@doi [\apj] {10.3847/1538-4357/aad091}, \href
  {http://adsabs.harvard.edu/abs/2018ApJ...863...85C} {863, 85}

\bibitem[\protect\citeauthoryear{{Chuard} et~al.,}{{Chuard}
  et~al.}{2018}]{chuard18}
{Chuard} D.,  et~al., 2018, \mn@doi [\aap] {10.1051/0004-6361/201731864}, \href
  {http://adsabs.harvard.edu/abs/2018A%26A...610A..34C} {610, A34}

\bibitem[\protect\citeauthoryear{{Churazov} et~al.,}{{Churazov}
  et~al.}{1993}]{1993ApJ...407..752C}
{Churazov} E.,  et~al., 1993, \mn@doi [\apj] {10.1086/172557}, \href
  {http://adsabs.harvard.edu/abs/1993ApJ...407..752C} {407, 752}

\bibitem[\protect\citeauthoryear{{Churazov}, {Khabibullin}, {Sunyaev}  \&
  {Ponti}}{{Churazov} et~al.}{2017a}]{2017MNRAS.465...45C}
{Churazov} E.,  {Khabibullin} I.,  {Sunyaev} R.,   {Ponti} G.,  2017a, \mn@doi
  [\mnras] {10.1093/mnras/stw2750}, \href
  {http://adsabs.harvard.edu/abs/2017MNRAS.465...45C} {465, 45}

\bibitem[\protect\citeauthoryear{{Churazov}, {Khabibullin}, {Sunyaev}  \&
  {Ponti}}{{Churazov} et~al.}{2017b}]{2017MNRAS.471.3293C}
{Churazov} E.,  {Khabibullin} I.,  {Sunyaev} R.,   {Ponti} G.,  2017b, \mn@doi
  [\mnras] {10.1093/mnras/stx1855}, \href
  {http://adsabs.harvard.edu/abs/2017MNRAS.471.3293C} {471, 3293}

\bibitem[\protect\citeauthoryear{{Clavel}, {Terrier}, {Goldwurm}, {Morris},
  {Ponti}, {Soldi}  \& {Trap}}{{Clavel} et~al.}{2013}]{clavel13}
{Clavel} M.,  {Terrier} R.,  {Goldwurm} A.,  {Morris} M.~R.,  {Ponti} G.,
  {Soldi} S.,   {Trap} G.,  2013, \mn@doi [\aap] {10.1051/0004-6361/201321667},
  \href {http://adsabs.harvard.edu/abs/2013A%26A...558A..32C} {558, A32}

\bibitem[\protect\citeauthoryear{{Clavel}, {Soldi}, {Terrier}, {Tatischeff},
  {Maurin}, {Ponti}, {Goldwurm}  \& {Decourchelle}}{{Clavel}
  et~al.}{2014}]{clavel14}
{Clavel} M.,  {Soldi} S.,  {Terrier} R.,  {Tatischeff} V.,  {Maurin} G.,
  {Ponti} G.,  {Goldwurm} A.,   {Decourchelle} A.,  2014, \mn@doi [\mnras]
  {10.1093/mnrasl/slu100}, \href
  {http://adsabs.harvard.edu/abs/2014MNRAS.443L.129C} {443, L129}

\bibitem[\protect\citeauthoryear{{Cotera}, {Erickson}, {Colgan}, {Simpson},
  {Allen}  \& {Burton}}{{Cotera} et~al.}{1996}]{cotera96}
{Cotera} A.~S.,  {Erickson} E.~F.,  {Colgan} S.~W.~J.,  {Simpson} J.~P.,
  {Allen} D.~A.,   {Burton} M.~G.,  1996, \mn@doi [\apj] {10.1086/177099},
  \href {http://adsabs.harvard.edu/abs/1996ApJ...461..750C} {461, 750}

\bibitem[\protect\citeauthoryear{{Dogiel}, {Chernyshov}, {Kiselev}  \&
  {Cheng}}{{Dogiel} et~al.}{2014}]{2014APh....54...33D}
{Dogiel} V.~A.,  {Chernyshov} D.~O.,  {Kiselev} A.~M.,   {Cheng} K.-S.,  2014,
  \mn@doi [Astroparticle Physics] {10.1016/j.astropartphys.2013.10.007}, \href
  {http://adsabs.harvard.edu/abs/2014APh....54...33D} {54, 33}

\bibitem[\protect\citeauthoryear{{Figer}, {Kim}, {Morris}, {Serabyn}, {Rich}
  \& {McLean}}{{Figer} et~al.}{1999}]{figer99}
{Figer} D.~F.,  {Kim} S.~S.,  {Morris} M.,  {Serabyn} E.,  {Rich} R.~M.,
  {McLean} I.~S.,  1999, \mn@doi [\apj] {10.1086/307937}, \href
  {http://adsabs.harvard.edu/abs/1999ApJ...525..750F} {525, 750}

\bibitem[\protect\citeauthoryear{{Figer} et~al.,}{{Figer}
  et~al.}{2002}]{figer02}
{Figer} D.~F.,  et~al., 2002, \mn@doi [\apj] {10.1086/344154}, \href
  {http://adsabs.harvard.edu/abs/2002ApJ...581..258F} {581, 258}

\bibitem[\protect\citeauthoryear{{Freeman}, {Doe}  \&
  {Siemiginowska}}{{Freeman} et~al.}{2001}]{sherpa}
{Freeman} P.,  {Doe} S.,   {Siemiginowska} A.,  2001, in {Starck} J.-L.,
  {Murtagh} F.~D.,  eds,  \procspie Vol. 4477, Astronomical Data Analysis. pp
  76--87 (\mn@eprint {} {astro-ph/0108426}), \mn@doi{10.1117/12.447161}

\bibitem[\protect\citeauthoryear{{Fruscione} et~al.,}{{Fruscione}
  et~al.}{2006}]{ciao}
{Fruscione} A.,  et~al., 2006, in Society of Photo-Optical Instrumentation
  Engineers (SPIE) Conference Series. p. 62701V, \mn@doi{10.1117/12.671760}

\bibitem[\protect\citeauthoryear{{Harrison} et~al.,}{{Harrison}
  et~al.}{2013}]{harrison}
{Harrison} F.~A.,  et~al., 2013, \mn@doi [\apj] {10.1088/0004-637X/770/2/103},
  \href {http://adsabs.harvard.edu/abs/2013ApJ...770..103H} {770, 103}

\bibitem[\protect\citeauthoryear{{Hong} et~al.,}{{Hong}
  et~al.}{2016}]{2016ApJ...825..132H}
{Hong} J.,  et~al., 2016, \mn@doi [\apj] {10.3847/0004-637X/825/2/132}, \href
  {http://adsabs.harvard.edu/abs/2016ApJ...825..132H} {825, 132}

\bibitem[\protect\citeauthoryear{{Koyama}, {Maeda}, {Sonobe}, {Takeshima},
  {Tanaka}  \& {Yamauchi}}{{Koyama} et~al.}{1996}]{1996PASJ...48..249K}
{Koyama} K.,  {Maeda} Y.,  {Sonobe} T.,  {Takeshima} T.,  {Tanaka} Y.,
  {Yamauchi} S.,  1996, \mn@doi [\pasj] {10.1093/pasj/48.2.249}, \href
  {http://adsabs.harvard.edu/abs/1996PASJ...48..249K} {48, 249}

\bibitem[\protect\citeauthoryear{{Krivonos} et~al.,}{{Krivonos}
  et~al.}{2014}]{K14}
{Krivonos} R.~A.,  et~al., 2014, \mn@doi [\apj] {10.1088/0004-637X/781/2/107},
  \href {http://adsabs.harvard.edu/abs/2014ApJ...781..107K} {781, 107}

\bibitem[\protect\citeauthoryear{{Krivonos} et~al.,}{{Krivonos}
  et~al.}{2017}]{K17}
{Krivonos} R.,  et~al., 2017, \mn@doi [\mnras] {10.1093/mnras/stx585}, \href
  {http://adsabs.harvard.edu/abs/2017MNRAS.468.2822K} {468, 2822}

\bibitem[\protect\citeauthoryear{Law \& Yusef-Zadeh}{Law \&
  Yusef-Zadeh}{2004}]{J174555}
Law C.,  Yusef-Zadeh F.,  2004, The Astrophysical Journal, 611, 858

\bibitem[\protect\citeauthoryear{{Madsen}, {Christensen}, {Craig}, {Forster},
  {Grefenstette}, {Harrison}, {Miyasaka}  \& {Rana}}{{Madsen}
  et~al.}{2017}]{straylight}
{Madsen} K.~K.,  {Christensen} F.~E.,  {Craig} W.~W.,  {Forster} K.~W.,
  {Grefenstette} B.~W.,  {Harrison} F.~A.,  {Miyasaka} H.,   {Rana} V.,  2017,
  \mn@doi [Journal of Astronomical Telescopes, Instruments, and Systems]
  {10.1117/1.JATIS.3.4.044003}, 3, 3

\bibitem[\protect\citeauthoryear{{Mori} et~al.,}{{Mori}
  et~al.}{2015}]{2015ApJ...814...94M}
{Mori} K.,  et~al., 2015, \mn@doi [\apj] {10.1088/0004-637X/814/2/94}, \href
  {https://ui.adsabs.harvard.edu/#abs/2015ApJ...814...94M} {814, 94}

\bibitem[\protect\citeauthoryear{{Murakami}, {Koyama}, {Sakano}, {Tsujimoto}
  \& {Maeda}}{{Murakami} et~al.}{2000}]{2000ApJ...534..283M}
{Murakami} H.,  {Koyama} K.,  {Sakano} M.,  {Tsujimoto} M.,   {Maeda} Y.,
  2000, \mn@doi [\apj] {10.1086/308717}, \href
  {http://adsabs.harvard.edu/abs/2000ApJ...534..283M} {534, 283}

\bibitem[\protect\citeauthoryear{{Ponti}, {Terrier}, {Goldwurm}, {Belanger}  \&
  {Trap}}{{Ponti} et~al.}{2010}]{2010ApJ...714..732P}
{Ponti} G.,  {Terrier} R.,  {Goldwurm} A.,  {Belanger} G.,   {Trap} G.,  2010,
  \mn@doi [\apj] {10.1088/0004-637X/714/1/732}, \href
  {http://adsabs.harvard.edu/abs/2010ApJ...714..732P} {714, 732}

\bibitem[\protect\citeauthoryear{{Ponti}, {Morris}, {Terrier}  \&
  {Goldwurm}}{{Ponti} et~al.}{2013}]{2013ASSP...34..331P}
{Ponti} G.,  {Morris} M.~R.,  {Terrier} R.,   {Goldwurm} A.,  2013, in {Torres}
  D.~F.,  {Reimer} O.,  eds,  Astrophysics and Space Science Proceedings Vol.
  34, Cosmic Rays in Star-Forming Environments. p.~331 (\mn@eprint {arXiv}
  {1210.3034}), \mn@doi{10.1007/978-3-642-35410-6_26}

\bibitem[\protect\citeauthoryear{{Revnivtsev} et~al.,}{{Revnivtsev}
  et~al.}{2004}]{2004A&A...425L..49R}
{Revnivtsev} M.~G.,  et~al., 2004, \mn@doi [\aap]
  {10.1051/0004-6361:200400064}, \href
  {http://adsabs.harvard.edu/abs/2004A%26A...425L..49R} {425, L49}

\bibitem[\protect\citeauthoryear{{Ross} \& {Fabian}}{{Ross} \&
  {Fabian}}{2005}]{2005MNRAS.358..211R}
{Ross} R.~R.,  {Fabian} A.~C.,  2005, \mn@doi [\mnras]
  {10.1111/j.1365-2966.2005.08797.x}, \href
  {http://adsabs.harvard.edu/abs/2005MNRAS.358..211R} {358, 211}

\bibitem[\protect\citeauthoryear{{Ryu}, {Nobukawa}, {Nakashima}, {Tsuru},
  {Koyama}  \& {Uchiyama}}{{Ryu} et~al.}{2013}]{2013PASJ...65...33R}
{Ryu} S.~G.,  {Nobukawa} M.,  {Nakashima} S.,  {Tsuru} T.~G.,  {Koyama} K.,
  {Uchiyama} H.,  2013, \mn@doi [\pasj] {10.1093/pasj/65.2.33}, \href
  {http://adsabs.harvard.edu/abs/2013PASJ...65...33R} {65, 33}

\bibitem[\protect\citeauthoryear{{Serabyn}, {Shupe}  \& {Figer}}{{Serabyn}
  et~al.}{1998}]{serabyn98}
{Serabyn} E.,  {Shupe} D.,   {Figer} D.~F.,  1998, \mn@doi [\nat]
  {10.1038/28799}, \href {http://adsabs.harvard.edu/abs/1998Natur.394..448S}
  {394, 448}

\bibitem[\protect\citeauthoryear{{Sunyaev} \& {Churazov}}{{Sunyaev} \&
  {Churazov}}{1998}]{1998MNRAS.297.1279S}
{Sunyaev} R.,  {Churazov} E.,  1998, \mn@doi [\mnras]
  {10.1046/j.1365-8711.1998.01684.x}, \href
  {http://adsabs.harvard.edu/abs/1998MNRAS.297.1279S} {297, 1279}

\bibitem[\protect\citeauthoryear{{Sunyaev}, {Markevitch}  \&
  {Pavlinsky}}{{Sunyaev} et~al.}{1993}]{1993ApJ...407..606S}
{Sunyaev} R.~A.,  {Markevitch} M.,   {Pavlinsky} M.,  1993, \mn@doi [\apj]
  {10.1086/172542}, \href {http://adsabs.harvard.edu/abs/1993ApJ...407..606S}
  {407, 606}

\bibitem[\protect\citeauthoryear{{Tatischeff}, {Decourchelle}  \&
  {Maurin}}{{Tatischeff} et~al.}{2012}]{T12}
{Tatischeff} V.,  {Decourchelle} A.,   {Maurin} G.,  2012, \mn@doi [\aap]
  {10.1051/0004-6361/201219016}, \href
  {http://adsabs.harvard.edu/abs/2012A%26A...546A..88T} {546, A88}

\bibitem[\protect\citeauthoryear{{Terrier} et~al.,}{{Terrier}
  et~al.}{2010}]{2010ApJ...719..143T}
{Terrier} R.,  et~al., 2010, \mn@doi [\apj] {10.1088/0004-637X/719/1/143},
  \href {http://adsabs.harvard.edu/abs/2010ApJ...719..143T} {719, 143}

\bibitem[\protect\citeauthoryear{{Terrier}, {Clavel}, {Soldi}, {Goldwurm},
  {Ponti}, {Morris}  \& {Chuard}}{{Terrier} et~al.}{2018}]{2018A&A...612A.102T}
{Terrier} R.,  {Clavel} M.,  {Soldi} S.,  {Goldwurm} A.,  {Ponti} G.,  {Morris}
  M.~R.,   {Chuard} D.,  2018, \mn@doi [\aap] {10.1051/0004-6361/201730837},
  \href {https://ui.adsabs.harvard.edu/#abs/2018A&A...612A.102T} {612, A102}

\bibitem[\protect\citeauthoryear{{Tsujimoto}, {Hyodo}  \& {Koyama}}{{Tsujimoto}
  et~al.}{2007}]{tsujimoto}
{Tsujimoto} M.,  {Hyodo} Y.,   {Koyama} K.,  2007, \mn@doi [\pasj]
  {10.1093/pasj/59.sp1.S229}, \href
  {http://adsabs.harvard.edu/abs/2007PASJ...59S.229T} {59, 229}

\bibitem[\protect\citeauthoryear{{Wang}, {Dong}  \& {Lang}}{{Wang}
  et~al.}{2006}]{wang}
{Wang} Q.~D.,  {Dong} H.,   {Lang} C.,  2006, \mn@doi [\mnras]
  {10.1111/j.1365-2966.2006.10656.x}, \href
  {http://adsabs.harvard.edu/abs/2006MNRAS.371...38W} {371, 38}

\bibitem[\protect\citeauthoryear{{Wenger} et~al.,}{{Wenger}
  et~al.}{2000}]{simbad}
{Wenger} M.,  et~al., 2000, \mn@doi [\aaps] {10.1051/aas:2000332}, \href
  {http://adsabs.harvard.edu/abs/2000A%26AS..143....9W} {143, 9}

\bibitem[\protect\citeauthoryear{{Yusef-Zadeh}, {Law}, {Wardle}, {Wang},
  {Fruscione}, {Lang}  \& {Cotera}}{{Yusef-Zadeh} et~al.}{2002}]{zadeh}
{Yusef-Zadeh} F.,  {Law} C.,  {Wardle} M.,  {Wang} Q.~D.,  {Fruscione} A.,
  {Lang} C.~C.,   {Cotera} A.,  2002, \mn@doi [\apj] {10.1086/340058}, \href
  {http://adsabs.harvard.edu/abs/2002ApJ...570..665Y} {570, 665}

\bibitem[\protect\citeauthoryear{{Zhang} et~al.,}{{Zhang}
  et~al.}{2015}]{2015ApJ...815..132Z}
{Zhang} S.,  et~al., 2015, \mn@doi [\apj] {10.1088/0004-637X/815/2/132}, \href
  {http://adsabs.harvard.edu/abs/2015ApJ...815..132Z} {815, 132}

\makeatother
\end{thebibliography}

\bsp	
\label{lastpage}
\end{document}